\documentclass{sig-alternate}
\input{complementary_newdefines.cls}
\usepackage{amssymb}
\usepackage{times}
\usepackage{amsfonts}

\usepackage{subfig}
\usepackage{float}
\usepackage{multirow, graphicx}
\usepackage{algorithm}
\usepackage{algpseudocode} 
\usepackage{amsmath}

\usepackage{listings}
\lstdefinestyle{TTSQL}{basicstyle=\ttfamily,
                        keywordstyle=\bfseries,
                        emphstyle=\itshape,
                        showstringspaces=true,
                        morekeywords={within}
                        }
\makeatletter
\newcommand{\lstuppercase}{\uppercase\expandafter{\expandafter\lst@token
                           \expandafter{\the\lst@token}}}
\newcommand{\lstlowercase}{\lowercase\expandafter{\expandafter\lst@token
                           \expandafter{\the\lst@token}}}
\makeatother

\usepackage{tabularx, multirow} 

\usepackage{simplemargins}
    \setleftmargin{0.65in}
    \setrightmargin{0.65in}
    \settopmargin{.65in}
    \setbottommargin{.65in}

\usepackage{setspace}

\usepackage{tikz}

\usepackage{xspace}
\usepackage{url}

\usepackage[breaklinks]{hyperref}
\usepackage{breakurl}
\usepackage{wrapfig}

\definecolor{mygreen}{rgb}{0,0.6,0}

\usepackage{collectbox}
\usepackage{enumitem}

\setlist[itemize]{leftmargin=*}
\usepackage{mdframed}
\mdfsetup{skipabove=0pt,skipbelow=0pt}

\makeatletter
\newcommand{\sqbox}{%
    \collectbox{%
        \@tempdima=\dimexpr\width-\totalheight\relax
        \ifdim\@tempdima<\z@
            \fbox{\hbox{\hspace{-.5\@tempdima}\BOXCONTENT\hspace{-.5\@tempdima}}}%
        \else
            \ht\collectedbox=\dimexpr\ht\collectedbox+.5\@tempdima\relax
            \dp\collectedbox=\dimexpr\dp\collectedbox+.5\@tempdima\relax
            \fbox{\BOXCONTENT}%
        \fi
    }%
}

\usepackage[most]{tcolorbox}

\DeclareMathAlphabet{\mathcal}{OMS}{cmsy}{m}{n}

\usepackage{tikz}
\newcommand{\ballnumber}[1]{\tikz[baseline=(myanchor.base)] \node[circle,fill=.,inner sep=1pt] (myanchor) {\color{-.}\bfseries\scriptsize #1};}
\usepackage{stackrel} 
\usepackage[none]{hyphenat}

\makeatother

\begin{document}
\title{{\Large \textsf{LPAD}}: Building Secure Data Storage in Enclave using Authenticated Log-Structured Merge Trees}



\title{Secure and Write-Efficient Storage in Enclave by {Authenticated Log-Structured Merge Trees}}

\title{
Authenticated Key-Value Stores with Hardware Enclaves
}

\conferenceinfo{venue}{year}
\numberofauthors{5} 
\author{
\alignauthor 
Yuzhe (Richard) Tang
\\       \affaddr{Syracuse University}
\\       \affaddr{Syracuse, NY USA}
\\       \email{ytang100@syr.edu}
\alignauthor 
Ju Chen
\\       \affaddr{Syracuse University}
\\       \affaddr{Syracuse, NY USA}
\\       \email{jchen133@syr.edu}
\alignauthor 
Kai Li
\\       \affaddr{Syracuse University}
\\       \affaddr{Syracuse, NY USA}
\\       \email{kli111@syr.edu}
\and
\alignauthor 
Jianliang Xu
\\       \affaddr{Hong Kong Baptist University}
\\       \affaddr{Kowloon Tong, Hong Kong}
\\       \email{xujl@comp.hkbu.edu.hk}
\alignauthor 
Qi Zhang
\\       \affaddr{IBM Research, TJ Watson}
\\       \affaddr{Yorktown Heights, NY USA}
\\       \email{Q.Zhang@ibm.com}
}
\date{}
\maketitle

\newtheorem{theorem}{Theorem}[section]
\newtheorem{lemma}[theorem]{Lemma}

\newtheorem{proposition}[theorem]{Proposition}
\newtheorem{invariant}[theorem]{Invariant}
\newtheorem{conjecture}[theorem]{Conjecture}
\newtheorem{corollary}[theorem]{Corollary}
\newtheorem{definition}[theorem]{Definition}

\newenvironment{example}[1][Example]{\begin{trivlist}
\item[\hskip \labelsep {\bfseries #1}]}{\end{trivlist}}
\newenvironment{remark}[1][Remark]{\begin{trivlist}
\item[\hskip \labelsep {\bfseries #1}]}{\end{trivlist}}

%
\newcommand{\wq}[1]{\footnote{\textcolor{blue}{(Wenqing: #1)}}}
\newcommand{\jc}[1]{\footnote{\textcolor{blue}{(Ju: #1)}}}

\providecommand{\swlog}{\texttt{swlog}\xspace}
\providecommand{\cwlog}{\texttt{cwlog}\xspace}
\providecommand{\pwlog}{\texttt{pwlog}\xspace}

\providecommand{\lsmpad}{{\sc LsmPAD}\xspace}
\providecommand{\trustkv}{{\sc LPAD-SGX}\xspace}
\providecommand{\vcompact}{{\sc EnclCompact}\xspace}
\providecommand{\mcompact}{{\sc MerkleCompact}\xspace}
\providecommand{\vlsm}{{\sc LSM$^2$}\xspace}
\providecommand{\lkvs}{{\sc LKVS}\xspace}

\providecommand{\att}{\textsf{Att}\xspace}
\providecommand{\readatt}{\textsf{retrieve}\xspace}
\providecommand{\writeatt}{\textsf{append}\xspace}

\providecommand{\enclaveread}{\textsf{readKV}\xspace}
\providecommand{\enclavewrite}{\textsf{writeKV}\xspace}

\providecommand{\EENTER}{\textsf{EENTER}\xspace}
\providecommand{\EEXIT}{\textsf{EEXIT}\xspace}
\providecommand{\ERESUME}{\textsf{ERESUME}\xspace}
\providecommand{\EINIT}{\textsf{EINIT}\xspace}
\providecommand{\ECREATE}{\textsf{ECREATE}\xspace}
\providecommand{\AEX}{\textsf{AEX}\xspace}
\providecommand{\EEXTEND}{\textsf{EEXTEND}\xspace}
\providecommand{\EADD}{\textsf{EADD}\xspace}
\providecommand{\EACCEPT}{\textsf{EACCEPT}\xspace}
\providecommand{\EAUG}{\textsf{EAUG}\xspace}

\providecommand{\vGet}{\textsf{vGet$_c$}\xspace}
\providecommand{\vPut}{\textsf{vPut$_c$}\xspace}
\providecommand{\dGet}{\textsf{vGet}\xspace}
\providecommand{\dPut}{\textsf{vPut}\xspace}

\providecommand{\agg}{\textsf{Agg}}
\providecommand{\komt}{\textsc{KMT}\xspace}
\providecommand{\komtone}{\textsc{KMT}$_1$\xspace}
\providecommand{\komttwo}{\textsc{KMT}$_2$\xspace}
\providecommand{\tomt}{\textsc{TMT}\xspace}

\providecommand{\vo}{\text{$\pi$}}
\providecommand{\authpath}{\textsf{AuthPath}}
\providecommand{\signbatch}{\textsf{Sign}}

\providecommand{\tableindex}{\sc \textsubscript{I}}
\providecommand{\tablebase}{\sc \textsubscript{B}}
\providecommand{\lkvDelete}{\textsf{Delete}}
\providecommand{\cppreput}{\textsf{prePut}}
\providecommand{\cppostget}{\textsf{postGet}}

\providecommand{\dbwrite}{\textsf{Write}\xspace}
\providecommand{\dbdelete}{\textsf{Delete}}
\providecommand{\dbread}{\textsf{Read}\xspace}
\providecommand{\lock}{\textsf{LockMap}}

\providecommand{\framework}{\textsc{MERKURY}\xspace}

\providecommand{\qmember}{\textsf{Membership}}

\providecommand{\elsm}{\textsf{eLSM}\xspace}
\providecommand{\elsmone}{\textsf{eLSM-P1}\xspace}
\providecommand{\elsmtwo}{\textsf{eLSM-P2}\xspace}

\providecommand{\lpad}{\textsf{LPAD}\xspace}
\providecommand{\lpadd}{\textsf{LPAD}}
\providecommand{\lpadmerge}{\textsc{Merge}\xspace}
\providecommand{\lpadsigmerge}{\textsc{DigMerge}\xspace}
\providecommand{\lsmmerge}{\textsc{Compaction}\xspace}
\providecommand{\numlevel}{{\relax\ifmmode{}q\else $q$\fi}\xspace}
\providecommand{\level}{{\relax\ifmmode{}L\else $L$\fi}\xspace}
\providecommand{\levelidx}{{\relax\ifmmode{}ii\else $ii$\fi}\xspace}
\providecommand{\ts}{{\relax\ifmmode{}ts\else $ts$\fi}\xspace}

\providecommand{\defeq}{\stackrel{\text{def}}{=}}

\providecommand{\scheme}{{\relax\ifmmode{}\Pi\else $\Pi$\fi}\xspace}
\providecommand{\gen}{\textsc{Gen}\xspace}
\providecommand{\enc}{\textsc{Enc}\xspace}
\providecommand{\dec}{\textsc{Dec}\xspace}
\providecommand{\keyvar}{{\relax\ifmmode{}K\else $K$\fi}\xspace}
\providecommand{\key}{{\relax\ifmmode{}k\else $k$\fi}\xspace}
\providecommand{\keyspace}{{\relax\ifmmode{}\mathbb{K}\else $\mathbb{K}$\fi}\xspace}
\providecommand{\secpar}{{\relax\ifmmode{}n\else $n$\fi}\xspace}
\providecommand{\mvar}{{\relax\ifmmode{}M\else $M$\fi}\xspace}
\providecommand{\m}{{\relax\ifmmode{}m\else $m$\fi}\xspace}
\providecommand{\msgspace}{{\relax\ifmmode{}\mathbb{M}\else $\mathbb{M}$\fi}\xspace}
\providecommand{\civar}{{\relax\ifmmode{}C\else $C$\fi}\xspace}
\providecommand{\ci}{{\relax\ifmmode{}c\else $c$\fi}\xspace}
\providecommand{\cispace}{{\relax\ifmmode{}\mathbb{C}\else $\mathbb{C}$\fi}\xspace}

\providecommand{\ind}{{\textsc{IND}\xspace}}

\providecommand{\eav}{{\textsc{EAV}\xspace}}
\providecommand{\eavmul}{{\textsc{EAV-MUL}\xspace}}
\providecommand{\indeav}{{\ind-\eav\xspace}}

\providecommand{\kpa}{{\textsc{KPA}\xspace}}
\providecommand{\indkpa}{{\ind-\kpa\xspace}}

\providecommand{\cpa}{{\textsc{CPA}\xspace}}
\providecommand{\oracle}{{\relax\ifmmode{}\mathcal{O}\else $\mathcal{O}$\fi}\xspace}
\providecommand{\oraclequery}{{\relax\ifmmode{}\mathcal{Q}\else $\mathcal{Q}$\fi}\xspace}
\providecommand{\indcpa}{{\ind-\cpa\xspace}}

\providecommand{\cca}{{\textsc{CCA}\xspace}}
\providecommand{\indcca}{{\ind-\cca\xspace}}

\providecommand{\gameprivenc}{\textsc{PrivK}\xspace}
\providecommand{\privke}{\textsc{PrivKE}\xspace}

\providecommand{\mac}{\textsc{MAC}\xspace}
\providecommand{\auth}{\textsc{Mac}\xspace}
\providecommand{\verify}{\textsc{Vrfy}\xspace}
\providecommand{\mactag}{{\relax\ifmmode{}{t}\else ${t}$\fi}\xspace}
\providecommand{\accept}{\textsc{1}\xspace}
\providecommand{\reject}{\textsc{0}\xspace}
\providecommand{\gamemac}{\textsc{MacForge}\xspace}
\providecommand{\cma}{\textsc{ACMA}\xspace}

\providecommand{\digest}{{\relax\ifmmode{}{d}\else ${d}$\fi}\xspace}
\providecommand{\dlength}{{\relax\ifmmode{}{l}\else ${d}$\fi}\xspace}
\providecommand{\msgblock}{{\relax\ifmmode{}{b}\else ${b}$\fi}\xspace}
\providecommand{\numblock}{{\relax\ifmmode{}{t}\else ${t}$\fi}\xspace}
\providecommand{\sizeblock}{{\relax\ifmmode{}{t}\else ${t}$\fi}\xspace}

\providecommand{\challenger}{{\relax\ifmmode{}\mathcal{C}\else $\mathcal{C}$\fi}\xspace}
\providecommand{\adversary}{{\relax\ifmmode{}\mathcal{A}\else $\mathcal{A}$\fi}\xspace}
\providecommand{\adv}{\textsc{Adv}\xspace}
\providecommand{\negl}{\textsc{negl}\xspace}

\providecommand{\de}{{\textsc{DE}\xspace}}
\providecommand{\probenc}{{\textsc{PE}\xspace}}
\providecommand{\ope}{{\textsc{OPE}\xspace}}

\providecommand{\expe}{\textsc{Exp}\xspace}
\providecommand{\var}{\textsc{Var}\xspace}
\providecommand{\entropy}{\textsc{Entropy}\xspace}
\providecommand{\coll}{\textsc{Coll}\xspace}

\providecommand{\randass}{{\relax\ifmmode{}\leftarrow\else $\leftarrow{}$\fi}\xspace}
\providecommand{\detass}{{\relax\ifmmode{}:=\else $:=$\fi}\xspace} 

\providecommand{\otp}{\textsc{OTP}\xspace}
\providecommand{\pad}{\textsc{Pad}\xspace}

\providecommand{\gn}{{\relax\ifmmode{}G\else $G$\fi}\xspace} 
\providecommand{\prg}{\textsc{PRG}\xspace}
\providecommand{\prf}{\textsc{PRF}\xspace}
\providecommand{\func}{{\relax\ifmmode{}f\else $f$\fi}\xspace} 
\providecommand{\funcspace}{{\relax\ifmmode{}\mathbb{F}\else $\mathbb{F}$\fi}\xspace} 
\providecommand{\funcgen}{{\relax\ifmmode{}{F}\else ${F}$\fi}\xspace} 
\providecommand{\prperm}{\textsc{PRP}\xspace}

\providecommand{\perm}{{\relax\ifmmode{}\pi\else $\pi$\fi}\xspace} 

\providecommand{\distinguisher}{{\relax\ifmmode{}\mathcal{D}\else $\mathcal{D}$\fi}\xspace} 

\providecommand{\sciinit}{\textsc{Init}\xspace}
\providecommand{\scigetbits}{\textsc{GetBits}\xspace}

\providecommand{\seed}{{\relax\ifmmode{}{s}\else ${s}$\fi}\xspace} 
\providecommand{\iv}{{\relax\ifmmode{}{IV}\else ${IV}$\fi}\xspace} 
\providecommand{\nonce}{{\relax\ifmmode{}{Nonce}\else ${Nonce}$\fi}\xspace} 

\providecommand{\protocol}{{\relax\ifmmode{}\Pi\else $\Pi$\fi}\xspace}
\providecommand{\trans}{{\textsc{Trans}\xspace}}
\providecommand{\gameke}{{\textsc{KE}\xspace}}

\providecommand{\fieldz}{{\relax\ifmmode{}\mathbb{Z}\else $\mathbb{Z}$\fi}\xspace}

\providecommand{\group}{{\relax\ifmmode{}\mathbb{G}\else $\mathbb{G}$\fi}\xspace}
\providecommand{\groupadd}{{\relax\ifmmode{}\mathbb{Z}\else $\mathbb{Z}$\fi}\xspace}
\providecommand{\groupmul}{{\relax\ifmmode{}\mathbb{Z}^*\else $\mathbb{Z}^*$\fi}\xspace}
\providecommand{\ggen}{{\relax\ifmmode{}g\else $g$\fi}\xspace}
\providecommand{\gorder}{{\relax\ifmmode{}q\else $q$\fi}\xspace}
\providecommand{\ginit}{{\relax\ifmmode{}\mathcal{G}\else $\mathcal{G}$\fi}\xspace}
\providecommand{\gmod}{{\relax\ifmmode{}N\else $N$\fi}\xspace}

\providecommand{\eulerfunc}{{\relax\ifmmode{}\psi\else $\psi$\fi}\xspace}
\providecommand{\resgroup}{{\relax\ifmmode{}\mathbb{G}\else $\mathbb{G}$\fi}\xspace}
\providecommand{\resgroupgen}{{\relax\ifmmode{ggenXXX}\else $ggenXXX$\fi}\xspace}
\providecommand{\p}{{\relax\ifmmode{}p\else $p$\fi}\xspace}
\providecommand{\q}{{\relax\ifmmode{}q\else $q$\fi}\xspace}
\providecommand{\gamepubk}{\textsc{PrivK}\xspace}

\providecommand{\gcd}{{\textsc{GCD}\xspace}}
\providecommand{\primenum}{{\relax\ifmmode{}p\else $p$\fi}\xspace}

\providecommand{\dlog}{{\textsc{DL }\xspace}}
\providecommand{\ddh}{{\textsc{DDH}\xspace}}
\providecommand{\cdh}{{\textsc{CDH}\xspace}}

\providecommand{\ke}{{\textsc{KeyExch }\xspace}}
\providecommand{\dhke}{{\textsc{DHKE}\xspace}}

\providecommand{\pki}{{\textsc{PKI }\xspace}}
\providecommand{\ca}{{\relax\ifmmode{}\mathcal{CA}\else $\mathcal{CA}$\fi}\xspace}
\providecommand{\cert}{{\textsc{Cert}}}
\providecommand{\server}{{\relax\ifmmode{}{S}\else ${S}$\fi}\xspace}
\providecommand{\client}{{\relax\ifmmode{}{C}\else ${C}$\fi}\xspace}

\providecommand{\pke}{{\textsc{PKE}}}
\providecommand{\pkey}{{\relax\ifmmode{}pk\else $pk$\fi}\xspace}
\providecommand{\skey}{{\relax\ifmmode{}sk\else $sk$\fi}\xspace}
\providecommand{\pemkey}{{\relax\ifmmode{}pemk\else $pemk$\fi}\xspace}

\providecommand{\sign}{\textsc{Sign}\xspace}

\providecommand{\gameauthenc}{\textsc{AuthK}\xspace}

\providecommand{\ecb}{\textsc{ECB}\xspace}
\providecommand{\cbc}{\textsc{CBC}\xspace}
\providecommand{\ctr}{\textsc{CTR}\xspace}

\providecommand{\ifeq}{{\relax\ifmmode{}\stackrel{?}{=}\else $\stackrel{?}{=}$\fi}\xspace}

\providecommand{\hashvar}{{\relax\ifmmode{}H\else $H$\fi}\xspace}
\providecommand{\hashfix}{{\relax\ifmmode{}h\else $h$\fi}\xspace}
\providecommand{\hashkey}{{\relax\ifmmode{}s\else $s$\fi}\xspace}
\providecommand{\gamecoll}{\textsc{HashColl}\xspace}
\providecommand{\crhash}{\textsc{CR-Hash}\xspace}
\providecommand{\varlen}{{\relax\ifmmode{}L\else $L$\fi}\xspace}
\providecommand{\fixlen}{{\relax\ifmmode{}l\else $l$\fi}\xspace}
\providecommand{\hmac}{\textsc{HMAC}\xspace}

\providecommand{\commit}{\textsc{Com}\xspace}
\providecommand{\comm}{{\relax\ifmmode{}comm\else $comm$\fi}\xspace}
\providecommand{\params}{{\relax\ifmmode{}params\else $params$\fi}\xspace}
\providecommand{\gamehiding}{\textsc{Hiding}\xspace}
\providecommand{\gamebinding}{\textsc{Binding}\xspace}

\providecommand{\mt}{{\relax\ifmmode{}\mathcal{MT}\else $\mathcal{MT}$\fi}\xspace}
\providecommand{\mdamgard}{{\relax\ifmmode{}\mathcal{MD}\else $\mathcal{MD}$\fi}\xspace}


\providecommand{\alice}{{\relax\ifmmode{}A\else $A$\fi}\xspace}
\providecommand{\bob}{{\relax\ifmmode{}B\else $B$\fi}\xspace}
\providecommand{\charlie}{{\relax\ifmmode{}C\else $C$\fi}\xspace}
\providecommand{\david}{{\relax\ifmmode{}D\else $D$\fi}\xspace}
\providecommand{\eva}{{\relax\ifmmode{}E\else $E$\fi}\xspace}
\providecommand{\fiona}{{\relax\ifmmode{}F\else $F$\fi}\xspace}

\providecommand{\packet}{{\relax\ifmmode{}pt\else $pt$\fi}\xspace}
\providecommand{\mitm}{\textsc{MitM}\xspace}

\providecommand{\tlshs}{\textsc{TLS-HS}\xspace}
\providecommand{\tlsrl}{\textsc{TLS-RL}\xspace}
\providecommand{\mix}{\textsc{MIX}\xspace}

\providecommand{\sig}{{\relax\ifmmode{}dig\else $dig$\fi}\xspace}

\providecommand{\predicate}{{\relax\ifmmode{}P\else $P$\fi}\xspace}
\providecommand{\answer}{{\relax\ifmmode{}a\else $a$\fi}\xspace}

\providecommand{\iproof}{{\relax\ifmmode{}\pi\else $\pi$\fi}\xspace}

\providecommand{\gameads}{\textsc{GameADS}\xspace}
\providecommand{\ads}{\textsc{ADS}\xspace}
\providecommand{\adsquery}{\textsc{Query}\xspace}
\providecommand{\adsupdate}{\textsc{Update}\xspace}
\providecommand{\adssetup}{\textsc{Setup}\xspace}
\providecommand{\adsrefresh}{\textsc{DigUpdate}\xspace}


\providecommand{\bdh}{\textsc{SBDH}\xspace}


\providecommand{\datakey}{{\relax\ifmmode{}k\else $k$\fi}\xspace}
\providecommand{\dataval}{{\relax\ifmmode{}v\else $v$\fi}\xspace}
\providecommand{\ts}{{\relax\ifmmode{}ts\else $ts$\fi}\xspace}

\providecommand{\lkvGet}{\textsc{Get}\xspace}
\providecommand{\lkvPut}{\textsc{Put}\xspace}
\providecommand{\lkvScan}{\textsc{Scan}\xspace}
\providecommand{\lkvCompact}{\textsc{Compact}\xspace}
\providecommand{\lkvFlush}{\textsc{Flush}\xspace}



\begin{abstract}
Authenticated data storage on an untrusted platform is an important computing paradigm for cloud applications ranging from data outsourcing, to cryptocurrency and general transparency logs. These modern applications increasingly feature update-intensive workloads, whereas existing authenticated data structures (ADSs) designed with in-place updates are inefficient to handle such workloads. In this paper, we address this issue and propose a novel authenticated log-structured merge tree (eLSM) based key-value store by leveraging Intel SGX enclaves.

We present a system design that runs the code of eLSM store inside enclave. To circumvent the limited enclave memory (128 MB with the latest Intel CPUs), we propose to place the memory buffer of the eLSM store outside the enclave and protect the buffer using a new authenticated data structure by digesting individual LSM-tree levels. We design protocols to support query authentication in data integrity, completeness (under range queries), and freshness. The proof in our protocol is made small by including only the Merkle proofs at selective levels. 

We implement eLSM on top of Google LevelDB and Facebook RocksDB with minimal code change and performance interference. We evaluate the performance of eLSM under the YCSB workload benchmark and show a performance advantage of up to $4.5X$ speedup.
\end{abstract}

\section{Introduction}
The practice of outsourcing data storage to the public cloud has been emerging and gaining popularity in security applications. For instance, using Amazon S3 to store the Bitcoin transactions (in cryptocurrency applications) or to host general transparency logs (e.g., in Google Certificate Transparency~\cite{me:ct,DBLP:conf/uss/MelaraBBFF15}) can significantly bring down the operational cost of these security-centric schemes and are adopted in practice~\cite{me:hostlog}. The modern security applications features user-generated content continuously generated in an intensive data-write stream (e.g., cryptocurrency transactions and user-requested certificates; see \S~\ref{sec:targetapp} for details).

Data authenticity is the primary concern for storing and serving data on the untrusted cloud services which are constantly caught compromised in the real world. To guarantee data authenticity, a common approach is to employ the protocols of authenticated data structures (ADSs) between an untrusted cloud server and trusted clients (i.e., data owner and query users).
However, existing ADS schemes~\cite{DBLP:conf/esa/Tamassia03,DBLP:journals/algorithmica/MartelNDGKS04,DBLP:conf/ccs/PapamanthouTT08,DBLP:conf/ccs/ZhangKP15,Devanbu03authenticdata,Pang:2004:AQR:977401.978163,DBLP:conf/sigmod/LiHKR06,DBLP:conf/vldb/PapadopoulosYP07,DBLP:conf/sigmod/YangPPK09,DBLP:journals/vldb/YangPPK09} have several major limitations. First, they are designed based on update-in-place structures (i.e., requiring excessive communication and large proofs between the data owner and the cloud for updating the ADS), leading to known inefficiency problems in handling data updates~\cite{DBLP:conf/eurocrypt/PapamanthouSTY13,DBLP:conf/ccs/QianZCP14}.
Second, existing schemes require the query users to verify the proof of results obtained from the cloud, which incurs high bandwidth and computation overheads at the query clients.

To address these limitations, in this paper, we leverage off-the-shelf hardware enclaves, in particular Intel Software Guard eXtension (SGX)~\cite{me:intelsgx}, and propose a novel authenticated key-value store based on LSM trees.
The motivation of our design is two-fold: 1. (Why LSM tree?) An LSM tree (log-structured merge tree) is a data structure that supports append-only writes and random-access data reads. Periodically, it conducts a batch operation, called \lsmmerge, that reorganizes the data layout for better read performance in the future. By this design, an LSM tree has performance advantages in serving high-speed write streams and is widely adopted as the external-memory index structure in many modern storage systems including Google's BigTable~\cite{DBLP:conf/osdi/ChangDGHWBCFG06}/LevelDB~\cite{me:leveldb}, Facebook RocksDB~\cite{me:rocksdb}, Apache HBase~\cite{me:hbase}, Apache Cassandra~\cite{me:cassandra} (see \S~\ref{sec:prelim:lsm} for details). 
2. (Why Intel SGX?) 
Without a trusted party, realizing an authenticated LSM tree on the cloud requires using costly cryptographic protocols, such as verifiable computations (VC~\cite{DBLP:journals/jacm/AroraLMSS98,DBLP:conf/sp/ParnoHG013,DBLP:conf/eurosys/SettyBVBPW13,DBLP:conf/sosp/BraunFRSBW13,DBLP:conf/ndss/WahbySRBW15}), to support verifiable \lsmmerge.\footnote{Another possible approach is to transfer the whole dataset back to trusted data owners over the Internet, which is also expensive.} With the advent of commercial trusted execution environments, notably Intel Software Guard eXtension or SGX~\cite{me:intelsgx}, it is possible to build a trusted execution environment or enclave in proximity to the untrusted cloud platform. This makes bulk data transfer feasible and should be promising to support verifiable and efficient \lsmmerge for authenticated LSM trees.
In addition, the query users can be alleviated from the burden of result verification. 

In our envisioned architecture, trusted cloud applications (e.g., a database server) run inside SGX enclaves and issue data read/write requests to our authenticated key-value store that is co-located in the cloud. 
In our system design, the code (namely the codebase of a vanilla LSM store\footnote{``LSM store'' denotes the class of key-value stores designed based on LSM trees.}) is placed inside the enclave, which relies on the Intel SGX SDK~\cite{me:sgxsdk} or an in-enclave Library OS (LibOS)~\cite{DBLP:journals/tocs/BaumannPH15,DBLP:conf/usenix/TsaiPV17} to handle system calls. In terms of placing data (e.g., program states), a naive design is to store the data in the memory region inside the enclave. When handling data of Gigabytes, this design, however, imposes huge memory pressure inside enclave and would cause significant performance slowdown. To be more specific, the current family of Intel CPUs support 128 MB physical memory in the enclave, and when the enclave memory hosts more than 128 MB, it causes expensive enclave paging~\cite{DBLP:conf/eurosys/OrenbachLMS17}. In our preliminary performance study, the overhead causes slowdown of more than two orders of magnitudes (See \S~\ref{sec:lpadp1perf} for details). More fundamentally, the slowdown is caused by the security needs to protect enclave memory. Even if Intel may remove the hard memory limits of 128 MB in future releases, putting the large data in a (larger) enclave still incurs the large slowdown by memory protection.

To circumvent the inefficiency, we propose to place the memory data outside the enclave. 
More precisely, among various memory data structures in an LSM store, we place the read buffer outside the enclave and leave other structures that often grow sublinearly with the data size inside the enclave, including index structures (e.g., a bloom filter), write buffer, etc. 
To ensure the integrity of the data placed outside the enclave, we propose an authenticated LSM tree, named by \elsm.
\elsm builds a forest of Merkle trees, each digesting a ``level'' in an LSM tree (see the preliminary of an LSM tree in \S~\ref{sec:prelim:lsm}). 
\elsm supports efficient reads and small-sized query proofs by presenting Merkle proofs at selective levels. We have proved the security of the query authentication schemes in \elsm (in \S~\ref{sec:security}).

We also build the \elsm systems on Google LevelDB~\cite{me:leveldb} and Facebook RocksDB~\cite{me:rocksdb}. 
In our systems, the \elsm Merkle proofs are embedded in individual data records in such a way that the proof of a query can be naturally constructed from the Merkle proofs embedded in the data records included in the query result. 
By this means, we minimize the code change needed in Google LevelDB, reducing performance interference at runtime. 
For RocksDB, 
the \elsm system is implemented as a middleware that does not require code change of the underlying RocksDB, but instead just relies on its callback interface~\cite{me:rocksdbfilter}. More specifically, we implement authenticated \lsmmerge in some event handlers in the \lsmmerge path of RocksDB. 
With this, we believe the add-on design of \elsm is generally applicable to any LSM stores.
By contrast, existing work in the field, notably Speicher~\cite{DBLP:conf/fast/BailleuTBFHV19}, all requires significant code change of underlying LSM store (see \S~\ref{sec:vsspeicher} for details). 

We conduct a comprehensive performance study of \elsm under the YCSB workload benchmark~\cite{DBLP:conf/cloud/CooperSTRS10}. 
The performance result shows that \elsm achieves lower operation latency than the baseline of update-in-place data structures by more than one order of magnitudes. Comparing with the \elsm design with memory buffers in the enclave, the memory placement outside the enclave achieves up to $4.5X$ speedup in most YCSB workloads.

The contributions made in this paper include the following:

1. This work addresses an emerging security workload, that is, supporting query authentication in the presence of frequent data updates. We propose a novel SGX-enabled authenticated key-value store.

2. We present the system designs of \elsm that are secure, efficient and generic. It places the memory data outside the enclave to circumvent the limited memory size in the enclave. It builds an authenticated LSM tree with small query proofs at selective tree levels. 
To the best of our knowledge, \elsm is the first data-authentication middleware on LSM stores, without any code change of the underlying store.

3. We implemented functional prototypes of \elsm on Google LevelDB and Facebook RocksDB. 
The code of our prototype is open-sourced~\cite{me:elsmcode}.
We conducted a comprehensive performance study under the YCSB workload benchmark that shows up to $4.5X$ performance advantage.


{\color{black}
\section{Preliminary: LSM Tree-based Storage Systems} 
\label{sec:prelim}
\label{sec:prelim:lsm}

This section presents the preliminaries of LSM trees and stores.
Readers who are familiar with the details of LSM stores may skip this section.
Additional background on Intel SGX is deferred to Appendix \S~\ref{sec:prel:sgx}.

{\bf Data structure}: A log-structured merge tree or LSM tree~\cite{DBLP:journals/acta/ONeilCGO96} is a data structure that organizes a dataset by so-called levels. A level is a collection of key-value records that are written in a time of proximity. Inside a level, it stores a sorted run of records, first ordered by data keys and then by time. Upon a write, an LSM tree stores it in the first level. The first level is thus a mutable data structure updated by individual data writes. Upon a read, it searches the tree by iterating through levels and finds the record that matches the queried data key. Periodically, normally at an offline time, the LSM tree runs a \lsmmerge operation that merges sorted runs in adjacent levels, to make room in lower levels\footnote{Levels in an LSM tree are indexed and the levels with smaller indexes are lower levels.} for upcoming writes and to facilitate reads in the future.
An LSM tree consists of totally $\numlevel$ levels, which are $\level_0, \level_1, ...$. 

{\bf Storage systems}: LSM trees have been recently adopted in the design of many modern storage systems, including Google BigTable~\cite{DBLP:conf/osdi/ChangDGHWBCFG06}/LevelDB~\cite{me:leveldb}, Apache HBase~\cite{me:hbase}, Apache Cassandra~\cite{me:cassandra}, Facebook RocksDB~\cite{me:rocksdb}, etc. 
In these systems, the LSM tree is adopted as an external-memory data structure that manages disk IOs. Specifically, the LSM tree buffers the first-level data in memory (stored in the MemTable structure) and, through the \lsmmerge operation, stores data at higher levels on disk. The data stored in MemTable is backed up by a log file on disk, named Write-Ahead Log or WAL. Data records are written to disk in a large data unit (e.g., several megabytes). Each data unit is persisted in a so-called SSTable file (Sorted Strings Table~\cite{DBLP:conf/osdi/ChangDGHWBCFG06,me:leveldb,me:rocksdb}). By this means, random-access writes are buffered in memory and the system causes only sequential writes to disk.

The LSM tree is also used as a primary index in the key-value store system. An SSTable is a file consisting of multiple data blocks. To support fast data reads in an SSTable, there is a compact B+ tree that indexes the different data blocks in the SSTable. In addition, a Bloom filter is built for each data block that indexes the records in the block. A Bloom filter can facilitate the case when the queried key is not found in the block.

In terms of performance, the LSM-tree based storage design represents a middle ground between the two classic designs, that is, the read-optimized update-in-place storage (e.g., B+ tree and many database indices~\cite{DBLP:books/bc/ElmasriN94,DBLP:books/daglib/0015084}) and the write-optimized log-structured storage (e.g., log-structured file systems~\cite{DBLP:books/kl/Rosenblum95}). 
On the one hand, an LSM tree (in an external memory model) serves data writes in an append-only fashion, in a way similar to log-structured file systems. On the other hand, it supports random-access reads without scanning the entire dataset, which is similar to update-in-place style B+ trees. An LSM tree reaps the benefits from both worlds, at the expense of assuming some offline hours to do the batched compaction operation.

\section{Research Formulation}
\label{sec:overview}

In this section, we present the research formulation by describing the motivating application and workloads, system model, security goals and design rationale of \elsm.

\subsection{Motivating Scenarios}
\label{sec:targetapp}

We present in depth a motivating scenario of our work, that is, general transparency. Other motivating applications, such as outsourced cloud databases, are explained in 
Appendix \S~\ref{sec:targetapp2}.

{\bf Application scenario}: General transparency becomes a popular design paradigm for building trusted computing systems~\cite{me:generaltransparency,me:generaltransparency2}, where internal events of a target system are recorded in and exposed to a publicly-auditable log. The transparency design is being adopted in many operational systems in the real world, including certificate transparency (or CT)~\cite{me:ct,me:ct2,me:ct3}, key transparency~\cite{DBLP:conf/uss/MelaraBBFF15}, and public Blockchains~\cite{me:bitcoin,me:eth}. In particular, CT exposes the certificates issued by known Certificate Authorities (CA) to the public in order for timely detection of mis-issued certificates~\cite{me:misissued1,me:misissued2}. The Blockchain records buyer-seller transactions in a public ledger and allows anyone to audit the transaction history for assurance of no invalid transactions.

\label{sec:workloads}

{\bf The application workload} features 1) an intensive stream of small-data writes and 2) data reads that retrieve a selective part of the log for auditing and cause random-access to the storage medium.
For instance, in Blockchain, transactions (i.e., small data writes) are continuously submitted by a large number of buyers/sellers and new Blockchain nodes who join the network as a simplified payment verification client (i.e., SPV client~\cite{me:spv}) randomly accesses part of the transaction history (i.e., random-access reads). In the case of CT log, certificate registration requests (i.e., small data writes) are continuously submitted by a large number of domain owners (in the low-cost CA schemes such as Let's Encrypt~\cite{me:letsencrypt}). Also, the log is randomly accessed by web browsers running TLS handshakes which want to validate the certificates being used.

The write-intensive workload excludes conventional database storage engines that are designed based on update-in-place data structures (e.g., B+ trees). In update-in-place structures, a data update needs to overwrite the previous version of the record at the exact location where the record is stored. An update incurs lookups and random-accesses of the record's previous location, leading to disk seeks and write amplification. 

LSM stores lend themselves to serving the write-intensive workload, thanks to the design of append-only writes. 
Due to this reason, many LSM stores are adopted in practice. 
For cryptocurrency and Blockchain, Google's LSM store LevelDB~\cite{me:leveldb} is widely adopted in the software stack of many Blockchain clients including Bitcoin core~\cite{me:ldbbitcoin}, Ethereum~\cite{me:ldbethereum}, HyperLedger Fabric~\cite{me:ldbhyperledger}, multichain/stream~\cite{me:ldbmultichain}, and other crypto-currencies.  LevelDB~\cite{me:leveldb} is used in Google CT systems~\cite{me:ct3} and RocksDB for the key transparency systems~\cite{me:kt,me:trillion,DBLP:conf/uss/MelaraBBFF15}.

However, in these real-world uses of LSM stores, the handling of data integrity is unsatisfactory. On the one hand, data integrity is a critical security property in our target applications.\footnote{By contrast, data confidentiality is less important in general transparency systems.}  For instance, in CT, returning a revoked certificate (violating data freshness) may connect a user to an impersonator, leading to security breaches. On the other hand, most existing transparency systems protect data integrity  by replicating data across many nodes. This design while decentralizes trusts incurs high overhead. Using SGX enclave is a much more lightweight approach. Thus, in this work, we are motivated to harden the data integrity in LSM stores by SGX enclaves.




\subsection{System Model}

In our system model, a trusted application runs inside the enclave on an untrusted cloud platform. The trusted application manages security-sensitive big-data and relies on a key-value store interface in enclave for data storage.
The trusted application emits a read/write workload with characteristic described above (i.e., write-intensive workloads). The interface between the trusted application and data storage is a standard one in key-value store, 
described in Equation~\ref{eqn:putget}. Given a key-value record $\langle{}\datakey,\dataval\rangle{}$, a write request is $\lkvPut(\datakey,\dataval)$ and a read is $\dataval=\lkvGet(\datakey)$. Here, we assume the enclave runs a timestamp manager that assigns to a read/write operation a unique timestamp reflecting the real time.

\begin{eqnarray}
\nonumber
\ts&=&\lkvPut(\datakey,\dataval)
\\
\nonumber
\langle{}\datakey,\dataval,\ts\rangle{}&=&\lkvGet(\datakey,\ts_q)
\\
\{\langle{}\datakey,\dataval,\ts\rangle{}\}&=&\lkvScan(\datakey_1,\datakey_2,\ts_q)
\label{eqn:putget}
\end{eqnarray}

The trusted application can run in multiple threads and issues the \lkvPut/\lkvGet operations concurrently.

\subsection{Security Goals}
\label{sec:security}
\label{sec:secdef}

The security goal of this work is primarily on query authenticity, while we also address data confidentiality in \S~\ref{sec:confidential} and \S~\ref{sec:elsmone}. The query authenticity describes that given a dataset, whether the result of a read reflects the latest state of the dataset. To formally describe our security goal, we present the threats and security definition below.

{\bf Threats}: The adversary in our model is the untrusted host outside the enclave. The adversarial host runs operating systems and the instance of LSM store. She can mount attacks to present forged query results to the enclave. Specifically, given a \lkvGet request, a malicious host can forge a fake result (breaking query-result integrity), or present a stale record (violating query freshness), or skip a legitimate record (violating query completeness). The definitions of these query-authenticity properties are described below.

{\bf Security definitions}: Given a read $\langle\datakey,\dataval,\ts\rangle=\lkvGet(\datakey,\ts_q)$, the query authenticity includes various correctness properties: 1) Query integrity is about whether the read-result $\langle{}\datakey,\dataval,\ts\rangle{}$ is a key-value record written by a legitimate write request before. If the read result is not written by the data owner, it violates the query integrity.
2) Query completeness is about whether a read result is complete. In case of point query (w.r.t. range query), the completeness is about the membership of a result and it prevents a legitimate record from being excluded in the result. For instance, if the store has a matching record to the read but it returns an empty result, the query completeness\footnote{We may use the terms query authenticity and membership alternatively.} is violated. 3) Query freshness states whether the result $\langle{}\datakey,\dataval,\ts\rangle{}$ has the largest timestamp (or is the latest) among all records of the queried key $\datakey$ and with a timestamp smaller/earlier than $\ts_q$.

\subsection{Design Motivation: Why LSM Tree based Digest Structure?}

To support query authentication (with membership), 
the conventional approach is building a single Merkle tree over the entire dataset and updating the Merkle tree  ``in place'' upon data updates. This update-in-place Merkle tree design is widely adopted in real-world systems such as digesting state in Ethereum~\cite{me:eth}. 

To serve write-intensive workloads, however, the update-in-place digesting approach would incur high performance overhead. Briefly, with small data records (e.g., tens or hundreds of bytes as in a ``tweets''), the hash digests which themselves are tens of bytes can grow comparable with the size of dataset itself. They need to be stored on a large storage medium such as disk. With digests stored on disk, the update-in-place digest structures cause random disk accesses and thus impose high overhead to the write path. 

In this work, we present a new digest structure based on an LSM tree, called \elsm digests, where the digests are updated in an append-only fashion. The motivation of this design is based on the well-known fact~\cite{DBLP:conf/sigmod/SearsR12,DBLP:journals/vldb/JermaineOY07,DBLP:conf/sosp/RajuKCA17,DBLP:journals/pvldb/AhmadK15} that the append-only design of an LSM tree leads to  performance advantages when serving write-intensive workloads. 
While existing LSM trees are applied to the data-access path in a storage system, our proposed \elsm presents a new paradigm that applies the LSM tree on the security path, particularly for efficient digest accesses.

\section{\elsmone: A Strawman Design}

In this section, we present a straightforward design 
of LSM store with SGX and name it by \elsmone. We then analyze the performance to motivate our primary design \elsmtwo presented in the next section.

\subsection{The \elsmone Design}
\label{sec:elsmone}

\begin{figure}[h!]
  \centering
  \includegraphics[width=0.5\textwidth]{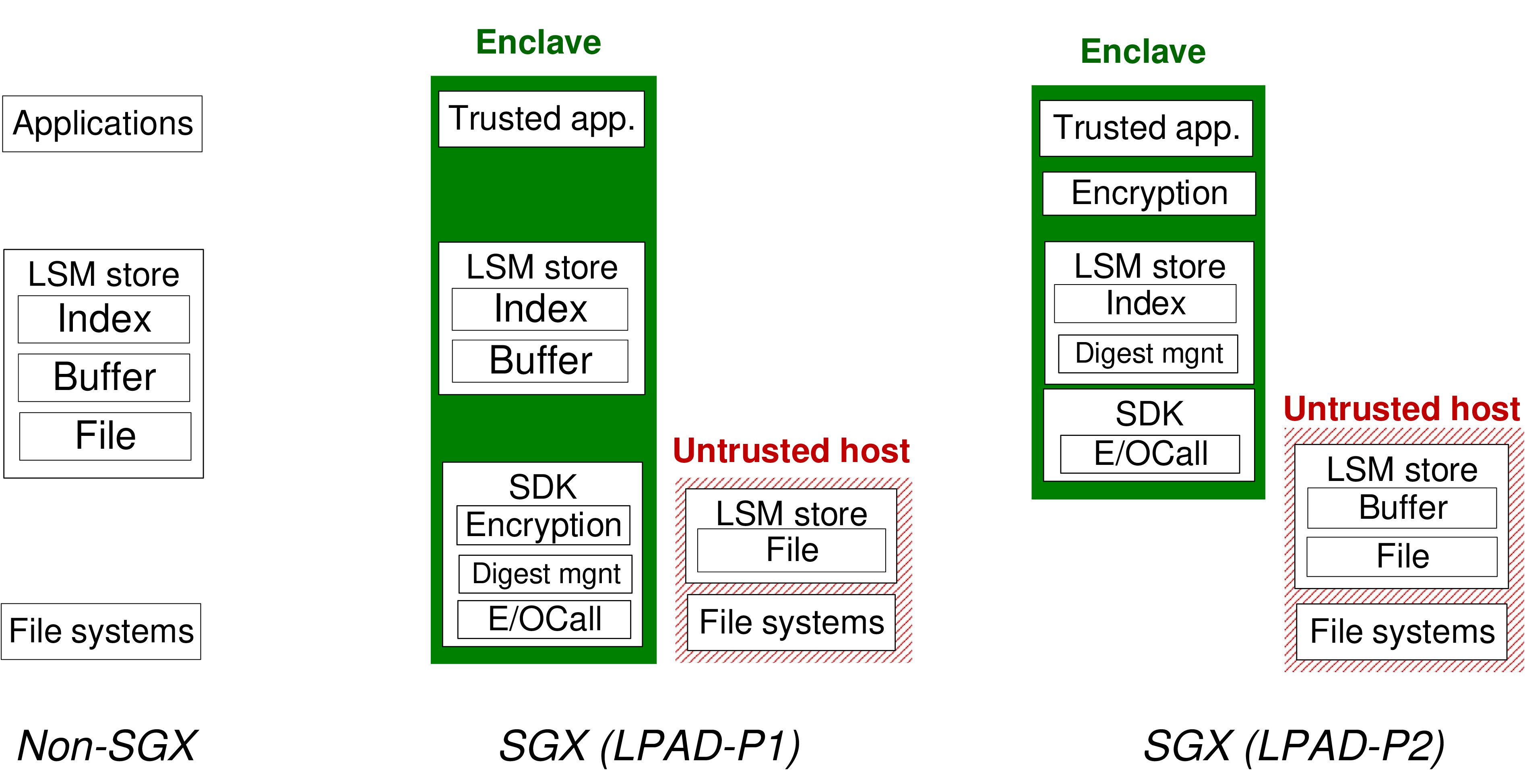}
  \caption{\elsm on SGX: System stack and designs}
  \label{fig:systemstack}
\end{figure}

{\bf Design space}: Recall that an LSM store is a user-space system that translates application-level data reads/writes to systems-level file reads/writes. To port an LSM store to SGX, in principle, one can place the code of the LSM store outside the enclave. That is, the LSM store runs in the untrusted world and the trusted application in enclave issues OCalls to send \lkvPut/\lkvGet requests to the store. However, this code-outside-enclave design requires excluding ``dynamically'' updated memory data from enclave and incurs design complexity for data authentication. We dismiss this alternative design 
(more details in Appendix \S~\ref{sec:codeplacement2}).
In this work, we focus on building an LSM store with code inside an SGX enclave.

{\bf A strawman}: We consider a strawman design, named \elsmone, that places the entire user-space codebase of an LSM store inside the enclave. The SSTable files are stored outside the enclave. The interaction between the enclave and the untrusted host occurs at the syscall levels, primarily for file management (e.g., \texttt{fwrite} and \texttt{fread}). Specifically, \elsmone places outside the enclave the files at all LSM-tree levels, including the WAL file at level $\level_0$ and SSTables at levels $\level_{\geq{1}}$. Inside the enclave, it runs the codebase of an LSM store and stores the intermediate data including indices and data buffers. The system architecture of \elsmone is depicted in Figure~\ref{fig:systemstack}.

We implement a functional system prototype of \elsmone 
on Google's LevelDB~\cite{me:leveldb}. To port Google LevelDB to SGX, we use Intel SGX SDK and modify the LevelDB codebase to call SDK's syscalls.

{\bf Security analysis}: \elsmone provides security on data confidentiality and authenticity through SGX SDK's data-protection mechanism. That is, SDK encrypts and digests the content of SSTable files stored outside the enclave, which ensures that an adversary host cannot extract information from the file content (in the ciphertext format) and forge a file block (whose integrity is verified by the hash proof of file block). Note that we do not address the confidentiality under side-channel attacks in SGX.

\subsection{Performance Analysis of \elsmone}
\label{sec:elsmtwo:motive}
\label{sec:lpadp1perf}

In this subsection, we first present a performance analysis of \elsmone that identifies performance problem serving a large dataset. We then use this observation to motivate our next design, \elsmtwo.

{\bf Observing \elsmone's performance}: We first focus on analyzing the read path of \elsmone. Recall that on the read path of an LSM store, the CPU accesses disk-resident data by buffering it in memory. \elsmone places the read buffers inside enclave, which may cause two performance problems: S1) The buffer in enclave incurs an extra data copy in the data-read path, that is, when the CPU accesses the data already buffered inside the untrusted memory, it creates a second data copy inside enclave. S2) When the in-enclave buffer grows large (e.g., beyond 128MB), it causes expensive enclave paging. For the common setting of disk-resident data, having a large read buffer is essential to the performance.

In order to quantify the performance slowdown caused by the in-enclave read buffer placement, we conduct a performance study based on our \elsmone implementation. For comparison, we also implement the placement of read buffer outside enclave. This is done by allocating the read buffer of LSM store in the untrusted memory. In the performance study, we store $5$ GB dataset (larger than untrusted memory) and drive a read-only workload that consists of reads of randomly distributed data keys. We vary the read buffer size $4$ MB to $2048$ MB. The read latency is reported in Figure~\ref{fig:eval:motive1}. 

\begin{figure}[!hbtp]
    \centering
    \includegraphics[width=0.35\textwidth]{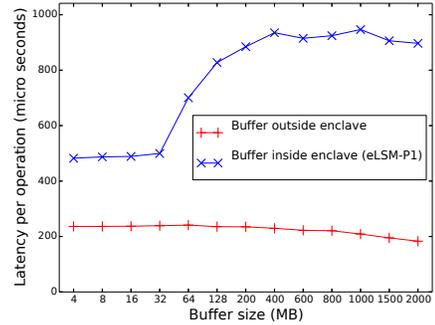}
    \caption{Placing read buffers inside/outside enclave and their performance difference}
    \label{fig:eval:motive1}
\end{figure}

The results show that when the data buffer is small, the in-enclave buffer incurs 2X latency of the out-enclave buffer. This performance difference is caused by the extra data copy in enclave (S1). When the data buffer is larger beyond 64 MB, the latency of accessing in-enclave buffer grows significantly and it incurs $4.5X$ latency. The performance slowdown of the large in-enclave buffer is due to the expensive enclave paging. This performance characteristic prompts us to place the memory buffers of an LSM store outside the enclave.


{\bf Motivating \elsmtwo Design}: The above performance observation suggests a favorable design by placing the read buffer of an LSM store outside the enclave. 
We propose \elsmtwo as a holistic system that materializes the design of placing read buffer outside the enclave. The \elsmtwo system features two placement strategies, that is, placing read buffer outside enclave and placing code and other memory (meta)data inside enclave.

Concretely, in \elsmtwo, the code of an LSM store runs inside the enclave, and it switches the execution out of the enclave only when serving system calls (e.g., file reads/writes). The enclave code accesses the data on the read path outside enclave, which includes read buffers and disk files. Other memory data is placed inside enclave, including the write buffer (at level $\level_0$) and file indices (at levels $\level_{\geq{1}}$). A common characteristic of data inside enclave is that they are meta-data in memory whose sizes are small enough (e.g., with  file indices of a couple of megabytes, it can accommodate millions of files) and thus they can be safely placed in enclave without causing enclave paging. 
\elsmtwo places only static memory data, that is, read buffers outside enclave, which significantly simplifies the complexity of a correct implementation.
Table~\ref{tab:lpad12} summarizes design choices made in \elsmone and \elsmtwo.

\section{\elsmtwo System}
\label{sec:elsmtwo:system}

\begin{table}[!htbp] 
\caption{Design choices of \elsmone and \elsmtwo}
\label{tab:lpad12}\centering{\scriptsize
\begin{tabularx}{0.5\textwidth}{ |c|X|X|X| }
  \hline
   & Code placement & Data placement & Digest structure \\ \hline
\elsmone
 (\S~\ref{sec:elsmone})
   & Inside enclave & Inside enclave & File granularity \\ \hline
\elsmtwo
 (\S~\ref{sec:elsmtwo:system})
   & Inside enclave & Outside enclave & Record granularity \\ \hline
\end{tabularx}
}
\end{table}

\subsection{System Overview}

Figure~\ref{fig:lpad1ex} depicts the system architecture of \elsmtwo which runs the code for operations \lkvPut,\lkvGet,\lsmmerge inside enclave. The memory data including the buffer at Level $\level_0$ and file indices at Levels $\level_{\geq{1}}$ are also placed inside the enclave.
The read buffers and all SSTable files at Levels $\level_{\geq{1}}$ are placed outside the enclave. The WAL file is also stored outside enclave. The figure also illustrates the dataset in the LSM store, which we will use throughout this section to describe the details of \elsmtwo system. In this example dataset, there is an LSM tree of three levels and six key-value records. Level $\level_1$ contains record $\langle{}A,9\rangle{}$, level $\level_2$ contains three records $\langle{}T,4\rangle{},\langle{}Z,7\rangle{},\langle{}Z,6\rangle{}$ and level $\level_3$ contains four records $\langle{}A,2\rangle{}$, $\langle{}T,0\rangle{}$, $\langle{}Y,3\rangle{}$, $\langle{}Z,1\rangle{}$. Here, we show the key-value record by its data key and timestamp. Record $\langle{}T,0\rangle{}$ is of data key $T$ and timestamp $0$, which is the oldest record. For simplicity, the data value is omitted in this example.

\begin{figure*}[!bhtp]
  \begin{center}
    \subfloat[System architecture]{%
       \includegraphics[width=0.35\textwidth]{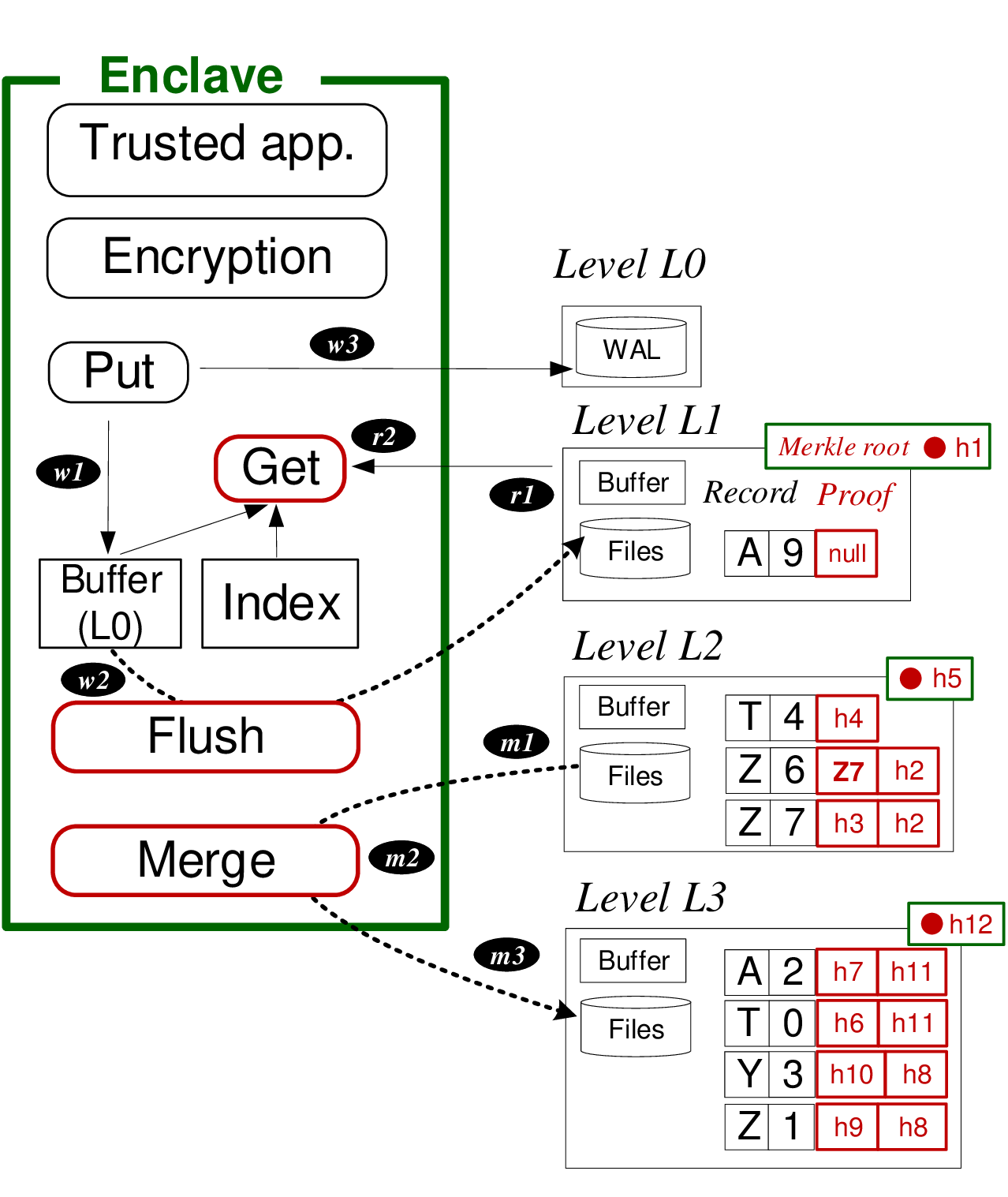}%
\label{fig:lpad1ex}
    }
 \hspace{0.05in}
    \subfloat[Authenticated \lsmmerge with Merkle trees: It merges two SSTables respectively at Levels $\level_2$ and $\level_3$ into two SSTables at Level $\level_3$. Each SSTable file is represented by a gray box.]{%
       \includegraphics[width=0.6\textwidth]{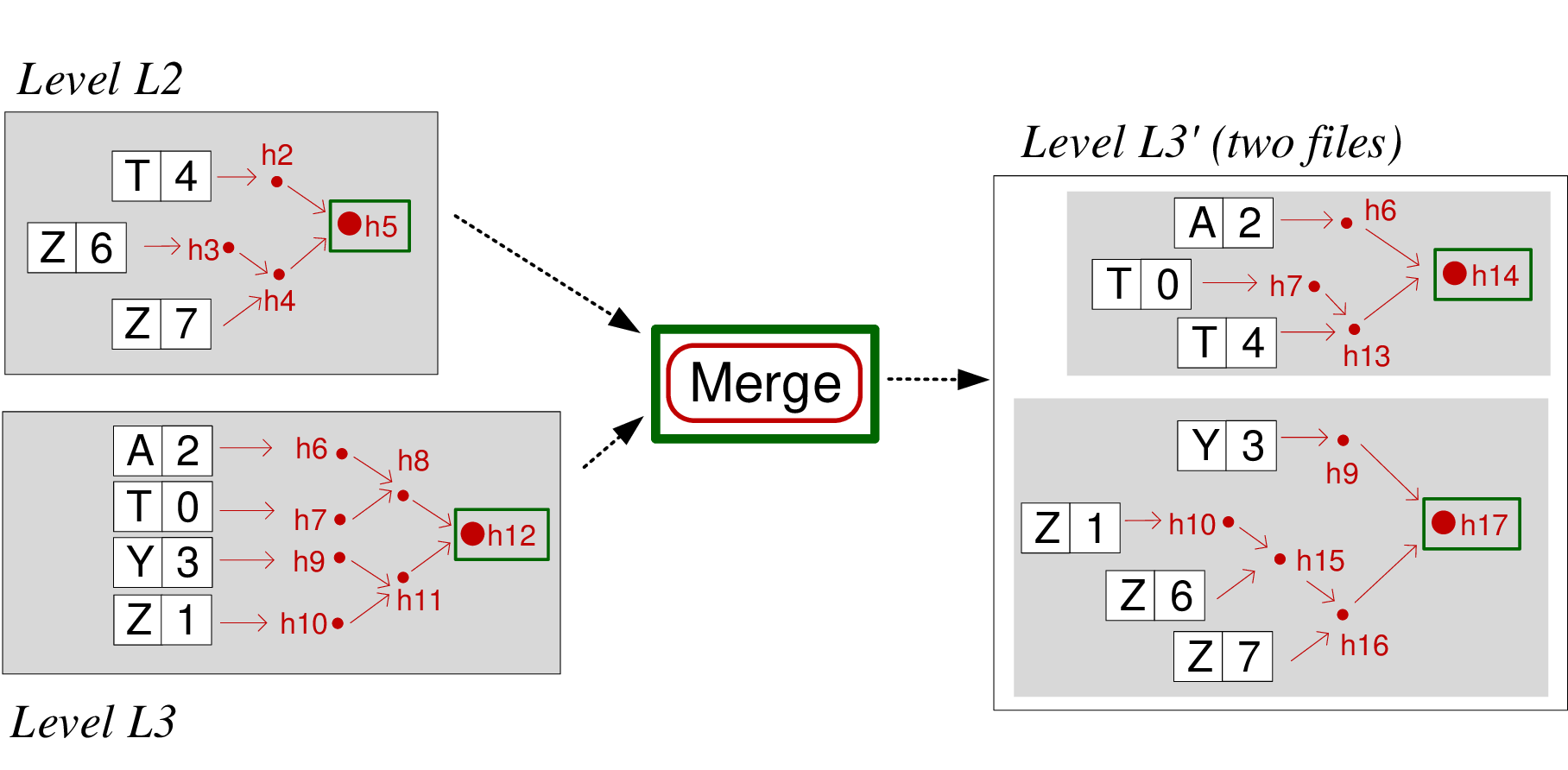}%
\label{fig:mergeex}
    }
  \end{center}
\caption{\elsmtwo system with an LSM tree of three levels: 
A rectangle depicts data and a rounded rectangle depicts code. Shapes in red are where \elsmtwo makes code change over the original LSM store. The root hashes (red dots in green boxes) are maintained with copies inside enclave.}
  \label{fig:lsmmodel}
\end{figure*}




{\bf Protecting data outside enclave}: Because \elsmtwo places outside enclave the data at non-zero levels, it entails data protection mechanisms. For data confidentiality, we require the data key in each record to be encrypted with deterministic encryption (DE), such that it can directly search the domain of ciphertext. We discuss the details of data confidentiality in \S~\ref{sec:confidential}. For data authenticity, we build the \elsm digest structure to authenticate the data outside enclave, as described next.

\subsection{Digest Structure}

To digest an LSM tree, we propose a novel authenticated data structure, \elsmtwo digests. There are two key designs: 1) \elsmtwo builds a ``forest'' of Merkle trees, each digesting one LSM-tree level and each having its root stored in the enclave. 2) In a per-level Merkle tree, data records of the same key are digested in hash chains and records of different keys are digested in a Merkle tree. In particular, the hash chain is built in a temporal order where the chain header is the oldest record and the tail is the newest record. In Figure~\ref{fig:lpad1ex}, each of the three LSM tree levels is associated with a Merkle tree. Case 1): For a level of distinct data keys, such as level $\level_3$ in Figure~\ref{fig:lpad1ex}, it builds the leaf set of the Merkle tree directly on the data records. For instance, $h_7=\hashvar(\langle{}A,2\rangle{})$ (\hashvar is a standard cryptographic hash algorithm with variable-length input) and $h_6=\hashvar(\langle{}T,0\rangle{})$. An intermediate node is the hash digest of the concatenation of the two children, for instance, $h_8=\hashvar(h_6\|h_7)$. Case 2) For a level that contains some records of the same keys, it constructs a hash chain over these records. For instance, level $\level_2$ contains two records of the same key, $\langle{}Z,7\rangle{}$ and $\langle{}Z,6\rangle{}$. \elsmtwo builds a \emph{hash chain} on these two records, that is, $h_4=\hashvar(\langle{}Z,7\rangle{}\|\hashvar(\langle{}Z,6\rangle{}))$. Then, it builds the Merkle tree over $h_4$ (for records of key $Z$) and $h_2$ (for record $\langle{}T,4\rangle$) for level $\level_2$.

To materialize the \elsmtwo digest structure, we present a simple storage design: Given a level $\level_i$ and its Merkle tree, each record at the level $\langle{}\datakey,\dataval\rangle$ is augmented with its \elsmtwo proof \iproof, that is, $\langle{}\datakey,\dataval\|\iproof_i\rangle{}$
Given a record, an \elsmtwo proof is the set of Merkle tree nodes (or hashes) that surround the path from the leaf node of the record to the root node.
For instance, in Figure~\ref{fig:lpad1ex}, the \elsmtwo proof for record $\langle{}A,2\rangle{}$ consists of hashes $h_7$ and $h_{11}$ (which are the siblings to nodes $h_6$ and $h_8$).

\subsection{Read/Write Protocol}
\label{sec:readwrite}
{\bf 
Data read path} starts with the trusted application issuing a read operation, $\dataval=\lkvGet(\datakey)$. 
The enclave looks up its index to locate the target level and file, and it then notifies the \elsmtwo store.
\ballnumber{r1}
The untrusted store serves the read operation on the target file and, in addition, runs algorithm $\iproof,\dataval=\adsquery_{\lkvGet}(\m,\datakey)$) to prepare a proof for authenticating the read result.
The proof consists of Merkle authentication paths or Merkle proofs~\cite{DBLP:conf/sp/Merkle80} at ``relevant'' LSM tree levels. 
Recall that a Merkle authentication path consists of the hashes surrounding the path from a leaf to the root in a Merkle tree and it can be used to verify the membership and non-membership of a record in a dataset.
\ballnumber{r2}
 The \elsmtwo store in the untrusted host then sends the result of \lkvGet(\datakey) as well as the proofs (\iproof) to the enclave. The enclave verifies the authenticity of the read result, by running algorithm Yes$|$No$=\verify_{\lkvGet}(\iproof,\dataval)$). 
The verification algorithm iterates through relevant levels, and, for each level, verifies the membership/non-membership of the queried data key (\datakey) using the Merkle proof (in \iproof) and locally stored root hash.

A strawman of designing the \elsmtwo proof is to scan all levels to prepare a proof (in algorithm \adsquery).
We propose to reduce the proof size by including only Merkle proofs of the levels no higher than the level of the result record.
This will allow algorithm \adsquery to stop early when it reaches the first level, say $\level_i$, that finds a matching record. The returned proof $\iproof=\iproof_1,...\iproof_{i}$, where $\iproof_1,...\iproof_{i-1}$ are the Merkle proofs for non-membership (there is not any matching record at levels $\level_1,\level_2,...\level_{i-1}$). $\iproof_i$ is the Merkle proof for membership (there is a matching record at level $\level_i$). All Merkle proofs after level $\level_i$, as will be seen, do not contain fresher records and are deliberately omitted. When there is no matching record, $i=\numlevel$.

{\it An example}:
In Figure~\ref{fig:lpad1ex}, suppose the trusted application issues $\lkvGet(Z)$ over dataset \m. In step \ballnumber{r1}, the untrusted host serves $\adsquery_{\lkvGet}(\m,Z)$ with authentic result $\langle{}Z,7\rangle{}$ (the benign case). $\langle{}Z,7\rangle{}$ is the newest record matching queried key $Z$ and is located at level $\level_2$. The proof is two Merkle authentication paths at levels $\level_1$ and $\level_2$. Note that there is no need to include level $\level_3$ in the \elsmtwo proof. Concretely, the proof at the first level is $\langle{}A,9\rangle{}$ (denoted by $\iproof_1$). The proof at the second level is $h_3,h_2$ (denoted by $\iproof_2$). Then in step \ballnumber{r2}, the enclave can verify the result authenticity in freshness and completeness based on the proof $\iproof=[\iproof_1,\iproof_2]$  (i.e., algorithm $\verify([\iproof_1,\iproof_2],\langle{}Z,7\rangle{}, [h_1,h_5])$).

Concretely, with the first-level proof $\iproof_1=\langle{}A,9\rangle{}$, the enclave verifies result authenticity by checking $\hashvar(\iproof_1)\stackrel[]{?}{=}h_5$. If the \verify algorithm runs through, it authenticates the fact that record $\langle{}A,9\rangle{}$ is the only record at level $\level_1$. From this, it can be derived that level $\level_1$ does not contain any record of key $Z$ (i.e., the non-membership of a data key $Z$ at level $\level_1$). 
With the second-level proof $\iproof_2=h_3,h_2$, the enclave verifies by checking $\hashvar(h_2\|\hashvar(\langle{}Z,7\rangle{}\|h_3))\stackrel[]{?}{=}h_5$. If successful, it authenticates the fact that a) record $\langle{}Z,7\rangle{}$ is a valid record at level $\level_2$ (result integrity), b) record $\langle{}Z,7\rangle{}$ is the newest record with key $Z$ (result freshness). Fact b) is based on that there are no other records of key $Z$ in the proof $\iproof_2$. Based on these two proofs, one can establish that record $\langle{}Z,7\rangle{}$ is the newest record in the dataset \m. 

Consider the malicious case when the untrusted host can return a stale record, say $\langle{}Z,6\rangle{}$, to the enclave. In this case, the malicious host can only present the following as a valid level-$\level_2$ proof, that is, $\iproof_2'=\langle{}Z,7\rangle{},h_2$. By this means, the enclave can verify the result integrity successfully by checking $\hashvar(h_2\|\hashvar(\langle{}Z,7\rangle{}\|\hashvar(\langle{}Z,6\rangle{})))\stackrel[]{?}{=}h_5$. However, as the newer result record $\langle{}Z,7\rangle$ has to be included in the proof $\iproof_2$, the enclave can detect that $\langle{}Z,6\rangle{}$ is not the most fresh record (violating freshness).

{\bf Data write path} starts with the trusted application issuing a write operation $\lkvPut(\datakey,\dataval)$.
To serve the write, the enclave maintains two in-enclave structures, a write buffer of level $\level_0$ and, for data recovery, a digest of the write-ahead log (WAL). Recall that a WAL stores recent data writes in temporal order and serves as the base to recover recent data in the case of fault. The storage of WAL is placed outside the enclave, while the enclave stores the hash digests of the WAL.

\ballnumber{w1}
Serving the write \lkvPut(\datakey,\dataval), the enclave first
assigns to the record to write the latest timestamp $\ts$. It then writes to the memory buffer of level $\level_0$ inside enclave. 
Serving a timestamped write $\lkvPut(\datakey,\dataval,\ts)$, the enclave iteratively update its WAL digest by $\sig'=H(\sig\|\langle{}\datakey,\dataval,\ts\rangle{})$. 
\ballnumber{w2}
When the write buffer at level $\level_0$ overflows, it is triggered 
to flush the content at Level $\level_0$ and to generate a file at Level $\level_1$. In the system of an LSM store, the codebase for flush is shared with that for \lsmmerge. 
\ballnumber{w3}
The enclave switches out to append the write to the WAL in the untrusted domain.
Enclave WAL can be extended to defend rollback attacks, which will be described in \S~\ref{sec:rollback}.

{\it An example}: 
In Figure~\ref{fig:lpad1ex}, suppose the application calls $\lkvPut(Y)$. The enclave assigns to the record the latest timestamp $10$. It updates the WAL digest from \sig to \sig', such that $\sig'=H(\sig\|\langle{}Y,10\rangle{}))$ (\ballnumber{w1}). 
The host appends the record to the WAL outside enclave (\ballnumber{w3}). 
If the buffer of Level $\level_0$ is overflown by the new record, it will sort all records stored in $\level_0$, and flush them to a new file at $\level_1$ (\ballnumber{w2}).

{\bf \lsmmerge path} starts with the trusted application in enclave issuing operation $(\level_i',\level_{i+1}')=\lsmmerge(\level_i,\level_{i+1})$. For simplicity, we consider the most basic form of \lsmmerge, namely, merging two adjacent levels. It is natural to extend it to more complicated cases such as merging more than two levels or merging subsets at the two levels. For the \lsmmerge across two levels, \elsmtwo carries out the computation inside enclave and only switches the execution outside enclave for file access. The process runs in the following steps: 
\ballnumber{m1} the enclave starts to issue OCalls to load all input files to untrusted memory so that the enclave can read the streams of data records (in their sorted order). \ballnumber{m2} 
The enclave then runs ``authenticated \lsmmerge'' that merges input data at the two levels into one level. Internally, the enclave needs to verify the authenticity of input data, to conduct the actual computation for \lsmmerge, to produce the digest of output data, and to generate the proofs embedded in the output data. We will describe in \S~\ref{sec:authmerge} the detailed system design of the authenticated \lsmmerge in enclave. 
\ballnumber{m3} 
The untrusted host makes effect of the \lsmmerge by flushing merged data and proof to disk. The enclave updates the per-level digests by the newly produced ones.

{\it An example}: 
In Figure~\ref{fig:lpad1ex}, 
suppose the application calls $\lsmmerge(\level_2,\level_3)$. In step \ballnumber{m1}, the host loads the data at the two levels from disk to memory (in the untrusted world).
In step \ballnumber{m2}, the enclave verifies the data authenticity of input levels by reconstructing the Merkle tree at level $\level_2$ (and $\level_3$) and by checking if its root hash is equal with $h_5$ (and $h_{12}$). It will then merge the two levels' data into one merged list, that is, from $\level_2=[\langle{}T,4\rangle{}, \langle{}Z,7\rangle{}, \langle{}Z,6\rangle{}]$ and $\level_3=[\langle{}A,2\rangle{}, \langle{}T,0\rangle{}, \langle{}Y,3\rangle{},\langle{}Z,1\rangle{}]$ to output level $\level_{3}'=[\langle{}A,2\rangle{}, \langle{}T,4\rangle{}, \langle{}T,0\rangle{}, \langle{}Y,3\rangle{}, \langle{}Z,7\rangle{}, \langle{}Z,6\rangle{}, \langle{}Z,1\rangle{}]$. Meanwhile, it builds the Merkle tree over the output list, and based on it, generates the proofs embedded in data records.
In step \ballnumber{m3}, the digest of the new Merkle tree replaces that of level $\level_3$ (i.e., $h_{12}$). $\level_2$ becomes an empty list and its digest is updated as well.

\subsubsection{Protocol Analysis}
\label{sec:security}
In this subsection we present the security analysis of the \elsmtwo protocol. We first define the record freshness as below:

\begin{definition} (Record freshness)
\label{def:freshness}
Given $\langle{}\datakey,\dataval,\ts\rangle{}=\lkvGet(\datakey,\m)$, \lkvGet returns a fresh result if the record $\langle{}\datakey,\dataval,\ts\rangle{}$ is the newest one (i.e., with the largest timestamp) in the dataset \m that matches data key $\datakey$.
\end{definition} 

The protocol correctness is defined as below:

\begin{definition} (Protocol correctness)
Given any state of dataset \m, if a \lkvGet returns a correct and fresh record w.r.t. state \m, the \verify algorithm will complete successfully (i.e., returning Yes). 
\end{definition} 

The protocol security is defined as below:

\begin{theorem} (Protocol security)
\label{thm:security}
Given a \lkvGet operation, if the enclave completes the \verify algorithm successfully, the \lkvGet result is correct and fresh (as in Definition~\ref{def:freshness}). 
\end{theorem}}


{\bf Security analysis}: 
We prove the protocol security by contradiction. Assume there exists an adversary who can forge a proof on a stale (yet correct)\footnote{The record integrity or correctness can be easily authenticated by the cryptographic hashes.} record on which the \verify algorithm completes successfully. Logically, there are two cases for a stale record: 1) The record is stale at the level that contains it, 2) The record is fresh at the level that contains it but there are more fresh records at other levels. In the following, we prove the contradiction that the \verify algorithm successfully completes on a stale record, respectively in the two cases.

For Case 1), suppose there is another record $\langle{}\datakey,\dataval',\ts'\rangle{}$ on the same level with result record $\langle{}\datakey,\dataval,\ts\rangle{}$ but fresher, that is, $\ts<\ts'$. The \lsmmerge in enclave guarantees\footnote{We assume program security in this work that the enclave program does not have exploitable bugs or run malware.} that the two records are sorted by timestamps (note that they are of the same data key) and the \elsmtwo builds a hash chain over them. More specifically,   $\hashvar(\langle{}\datakey,\dataval',\ts'\rangle{}\|...\hashvar(\langle{}\datakey,\dataval,\ts\rangle{}\|...))$. In other words, the fresh record $\langle{}\datakey,\dataval',\ts'\rangle{}$ is among the neighbors of the path from the stale record $\langle{}\datakey,\dataval,\ts\rangle{}$ to the Merkle root node (think for an example, $\langle{}Z,7\rangle{}$ is the neighbor of the path from $\langle{}Z,6\rangle{}$ to root as in Figure~\ref{fig:mergeex}). In order for the adversary to forge a proof that can be verified successfully, she either breaks the security of cryptographic hashes (1a) or has to include \emph{all} the neighbors of the path from itself to the root (1b). The former case (1a) is hard to any computationally bounded adversary (more formally, probabilistic polynomial time adversary). In the latter case (1b), the fresher record included in the neighbors is exposed to the enclave who will not pass the freshness check in the \verify algorithm as it can simply detect that the result record is not fresh (e.g., $\langle{}Z,7\rangle{}$ has a larger timestamp and is fresher than $\langle{}Z,6\rangle{}$). Therefore, the \verify algorithm cannot complete successfully on a stale record. A contradiction is found  in Case 1. 

For Case 2), suppose the result record $\langle{}\datakey,\dataval,\ts\rangle{}$ resides at level $\level_i$ and there exists a fresher record $\langle{}\datakey,\dataval',\ts'\rangle{}$ stored at another level $\level_j$. There are two cases: 2a) $j<i$ and 2b) $j>i$. For Case 2a), successfully passing \verify algorithm in \elsmtwo, it requires a non-membership proof of data key $\datakey$ at level $\level_j$. This contradicts the assumption that $\level_j$ does contain a record of key \datakey, namely $\langle{}\datakey,\dataval',\ts'\rangle{}$. For Case 2b), \elsmtwo does not require including any proof for the levels with index higher than $i$ (recall that given a \lkvGet result residing at level $\level_i$, an \elsmtwo proof includes Merkle proofs for levels $\level_1,\level_2,...\level_i$ but excludes the ones on any higher levels $\level_{i+1},\level_{i+2},...$). The following lemma guarantees that those levels higher than $\level_i$ (i.e., with larger index values) cannot include any records fresher than $\langle{}\datakey,\dataval,\ts\rangle{}$.

\begin{lemma}
\label{thm:temporalorder}
In \elsmtwo with the in-enclave \lsmmerge, a record residing at a lower level must have a larger timestamp than any record of the same data key at a higher level. That is, given any two records of the same data key, say $\langle{}\datakey,\dataval,\ts\rangle{}$ residing at level $\level_i$ and $\langle{}\datakey,\dataval',\ts'\rangle{}$ at level $\level_{i'}$, it holds that  $i<i'$ if and only if  $ts>ts'$.
\end{lemma}

For instance, in Figure~\ref{fig:lpad1ex}, at any level, say $\level_3$, records are sorted, from key $A$ to $T$ to $Z$. Lemma~\ref{thm:temporalorder} requires that in Figure~\ref{fig:lpad1ex}, an older record $A$ with timestamp $2$ is stored on a higher level $\level_3$ than the level a newer record $\langle{}A,9\rangle{}$ is stored (which is level $\level_1$). To prove Lemma~\ref{thm:temporalorder}, the key intuition is that \elsmtwo only allows moving key-value records from lower levels to higher levels, but not in the reversed order.

{\bf Meta-data authenticity}: In our implementation, meta-data including Bloom filters, file indices, etc. are placed inside enclaves. The metadata authenticity is protected by the enclave. 

\subsection{Range Query and Deletes}

{\bf Processing Range Query}: \elsmtwo supports range query processing with completeness. Given a queried key range, it iterates through \emph{all} levels (unlike the case of exact-match query) and for each level, the untrusted world presents the range-query proof from the Merkle tree at the level. 

Within one level, the range proof is constructed by treating the Merkle tree as a segment tree which is a classic data structure in computational geometry. A segment tree is essentially a full binary tree where each intermediate tree node represents a segment (or interval of data keys). Given a range query $L$, the segment tree can present $2\log{L}$ intermediate nodes, whose segments are union-ed to cover range $L$~\cite{DBLP:conf/iptps/ZhengSLS06}. Based on this view, \elsm constructs a range proof by the sibling nodes of the segments covering the queried range. 

For instance, suppose in Figure~\ref{fig:lsmmodel}, a range query is $\lkvScan([S,U])$ against Level $\level_3$. The query proof will include 1) records which fall in or enclose the range, that is, $\langle{}T,0\rangle{}, \langle{}Y,3\rangle{}, \langle{}Z,1\rangle{}$, 2) the range proof that authenticates these records, that is, $h_6$. At level $\level_3$, the segment union that covers the three records includes intermediate tree nodes $h_{11}$ and $h_7$. Their siblings are $h_6$ and $h_8$. While $h_8$ can be reconstructed from $\langle{}T,0\rangle{}$ and $h_6$, it can be omitted, leave the range proof to be $h_6$. In this case, a query verifier can authenticate the query completeness, as records $\langle{}T,0\rangle{}, \langle{}Y,3\rangle{}, \langle{}Z,1\rangle{}$ are authenticated and they consecutive in the leaf set of the Merkle tree. 

{\bf Security Analysis (Range Completeness)}: The range-query proof authenticates query completeness. Informally, the security can be derived from the following facts: 1) The hash functions used in Merkle tree are collision resistant. 2) The data records in a level are sorted by data keys. 3) The records in the query proof form a range of data keys that cover the range in the query.  

{\bf Handling Deletes}: \elsm relies on the underlying LSM store to support delete operations. Briefly, a delete operation stores a so-called tombstone record in the LSM store, which triggers the next \lsmmerge call to physically drop the records of the same key being merged. Before the \lsmmerge, a \lkvGet operation may return the tombstone record which is interpreted by \elsm as ``no record''.

\subsection{System Implementation}

We have implemented \elsmtwo on Google LevelDB~\cite{me:leveldb} and Facebook RocksDB~\cite{me:rocksdb}. For LevelDB, \elsm is implemented by directly changing the LevelDB codebase. This implementation approach has the advantage in performance. For RocksDB, \elsm is implemented as a RocksDB add-on, that is, without code changes to RocksDB. In this subsection, we present the LevelDB-based implementation in \S~\ref{sec:readpath},~\ref{sec:authmerge} and the RocksDB implementation in \S~\ref{sec:impl:rocksdb}.

In the protocol of \elsmtwo, a key-value record is stored with its proof, that is, $\langle{}\datakey,\dataval\|\iproof_i\rangle{}$, where $i$ is the index of the level where the record is currently located. 
To implement the embedded proof in an LSM store, it is required to add the code change in two paths, that is, a) the \lsmmerge paths for updating records' proof when they are merged to a different level (Note that the proof is sensitive to which level the record is located), and b) the \lkvGet path where the proof is used to authenticate the record membership/non-membership in a level. 
Next, we present the code change on the \lkvGet path in \S~\ref{sec:readpath}, and then present that on the \lsmmerge path in \S~\ref{sec:authmerge}.

\subsubsection{Query Verification in Read Path}
\label{sec:readpath}

The code change on the \lkvGet path serves two purposes: First, for the data stored in the untrusted world, \elsmtwo needs to present non-membership proofs for the levels that do not have a matching record. Instead of returning null for non-membership levels, \elsmtwo returns the two neighboring records whose keys are smaller and larger than the queried key. For instance, in Figure~\ref{fig:lpad1ex}, when it queries $\lkvGet(k=B)$ on level $\level_3$, it returns records $\langle{}A,2\rangle{}$ and $\langle{}T,0\rangle{}$ (with their \elsmtwo proofs). 

Second, inside the enclave, \elsmtwo requires implementing the \verify algorithm to verify the query authenticity given a proof. Note that the proof can be directly extracted from the result record returned from the untrusted part of \elsmtwo.

\label{sec:mmap}
{\bf Support \texttt{mmap} reads}: LevelDB supports the data reads in two ways: Read through a user-space read buffer and read through \texttt{mmap}'ed files. \elsm implementation over LevelDB supports both read paths in LevelDB. On the buffer-based read path, \elsmtwo allows the enclave code to access the buffer outside the enclave and let the untrusted code manage the buffer for eviction. 
On the \texttt{mmap}-based read path, it switches out of the enclave, upon opening a file, to \texttt{mmap} the file to the untrusted memory. Then the enclave code directly accesses the \texttt{mmap}'ed file in the untrusted memory.

\subsubsection{Authenticated \lsmmerge}
\label{sec:authmerge}

Recall that \elsmtwo runs the \lsmmerge computation inside the enclave by accessing data from the untrusted world. For data authenticity, the in-enclave \lsmmerge needs to be augmented with three extra steps: a) Before the \lsmmerge, the enclave authenticates the input data read from the untrusted world. b) After the \lsmmerge, the enclave needs to digest the output data and store the digest in the enclave. c) In the end, it generates from the output Merkle tree the proofs embedded in individual data records.

{\bf Integrating the \lsmmerge authentication} with the underlying LSM store (i.e., LevelDB and RocksDB) is realized by event handling. Concretely, the compaction process in an LSM store triggers various internal events. Two events are of particular interest to our implementation, that is, 
\texttt{Filter()} and \texttt{OnTableFileCreated()}. Event \texttt{Filter()} occurs whenever the compaction produces a data record. Event \texttt{OnTableFileCreated()} occurs whenever the compaction (or other procedure like flush) produces a new file on disk. In RocksDB, these two events are exposed in its callback functions (as will be described in details). In LevelDB, we have to modify the LevelDB codebase to expose the events (i.e., adding hooks) to applications. 
The pseudo-code that implements authenticated \lsmmerge by the two callbacks is depicted in Figure~\ref{lst:authmerge}. \elsmtwo runs the codebase of LSM store and the implemented callback functions in enclave. 
Steps a) and b): In \texttt{Filter()}, it constructs two types of Merkle trees, respectively for the input files and the output file. Given a key-value record, it parses the file name and level information from the value field and then updates the input Merkle tree at the corresponding level (Line 20 in Figure~\ref{lst:authmerge}). It also builds an output Merkle tree for the level the compaction output is located (Line 21). When the compaction finishes, it checks the equality of the Merkle root hash stored in enclave and the root hashes of the input Merkle trees reconstructed. If the equality check passes, the Merkle root hash for the output file takes effect (Line 36-39). 
Step c): In \texttt{OnTableFileCreated()}, it embeds the Merkle proof in the individual data records in the output file. 

{\bf Merkle tree construction}: Constructing a Merkle tree is a primitive operation in \elsmtwo, specifically for authenticating the input and output of a \lsmmerge. The Merkle tree in our scheme is constructed incrementally with the arrival of records in the input/output data stream. That is, when a data record arrives, the enclave checks if the record shares the data key with the  previous record. If this is the case, the enclave builds a hash chain on the previous record. Otherwise, it will incrementally build a Merkle tree. For instance, in Figure~\ref{fig:mergeex}, upon reading the output record $\langle{}T,4\rangle{}$, the enclave checks if the previous record (i.e., $\langle{}T,0\rangle{}$) has the same data key. Since the two records are of the same key, it then builds a hash chain over them. In the next iteration when the output record $\langle{}Y,3\rangle{}$ arrives, the previous record (i.e., $\langle{}T,4\rangle{}$) has a different data key, 
it then treats as a leaf of the Merkle tree the hash of previous record, namely $h_3$.

\definecolor{mygreen}{rgb}{0,0.6,0}
\begin{figure}[h!]
\lstset{ %
  backgroundcolor=\color{white},   
  basicstyle=\scriptsize\ttfamily,        
  breakatwhitespace=false,         
  breaklines=true,                 
  captionpos=b,                    
  commentstyle=\color{mygreen},    
  deletekeywords={...},            
  escapeinside={\%*}{*)},          
  extendedchars=true,              
  keepspaces=true,                 
  keywordstyle=\color{blue},       
  language=Java,                 
  morekeywords={*,...},            
  numbers=left,                    
  numbersep=5pt,                   
  numberstyle=\scriptsize\color{black}, 
  rulecolor=\color{black},         
  showspaces=false,                
  showstringspaces=false,          
  showtabs=false,                  
  stepnumber=1,                    
  stringstyle=\color{mymauve},     
  tabsize=2,                       
  title=\lstname,                  
  moredelim=[is][\bf]{*}{*},
}
\begin{lstlisting}
Record prev_r; Hash prev_h;
MHT[] input_mhts; MHT output_mht;

void MHT_add(Record r, MHT mht) {
    if (prev_r == NULL) h=H(r);
    else if (r.datakey != prev_r.datakey) {
        h=H(r);
        mht.add_leaf(prev_h);
    } else h=H(prev_h||r);
    prev_r = r; prev_h = h;}

void vanilla_compaction(SSTable La, SSTable Lb, 
           int filter(Record a)
           int onTableFileCreated(SSTable f));

void auth_filter(Record r) {
    MHT_add(r, input_mhts[parseFile(r)]);
    MHT_add(r, output_mht);}

int auth_onTableFileCreated(SSTable f)){
    SSTable f_new;
    for(record r : f){
        r.addEmbeddedProof(output_mht);
        f_new.write(r);}
    f = f_new;}

void auth_compaction(SSTable La, SSTable Lb){
    vanilla_compaction(La, Lb, 
            auth_filter, auth_onTableFileCreated);
    //check digest equality on input and then sign
    if(input_mhts[La].roothash == La.roothash &&
       input_mhts[Lb].roothash == Lb.roothash)
           Lb.roothash = output_mht.roothash;}
\end{lstlisting}
\vspace{-0.3in}
\caption{Authenticated \lsmmerge in \elsm}
\label{lst:authmerge}
\end{figure}

Our implementation retains the \lsmmerge features in the vanilla LSM store, including versioning policies, tombstone delete, etc. 

{\bf Multi-threading}: \elsmtwo supports concurrent operations in a multi-threaded enclave.
For concurrent reads/writes in MemTable at Level $\level_0$, \elsmtwo implementation relies on the synchronization support (i.e., mutex and condition variables) in enclave provided by Intel SGX SDK~\cite{me:sgxsync}. 
Concurrent reads on upper levels $\level_{\geq{1}}$ can be processed in parallel without synchronization.

Concurrent \lsmmerge with reads/writes needs to be synchronized, as is in LevelDB. For instance, when an SSTable produced by a \lsmmerge replaces an old SSTable at the same level, the replacement operation needs to be synchronized with any pending read on the old file.
This type of synchronization in \elsmtwo is realized by checks during query verification \verify, such that a read before/after the file replacement can only be verified successfully with the Merkle hash of the old/new SSTable file.
To implement this, the file replacement outside enclave synchronously calls into the enclave (through a blocking ECall) in order to update the file hash maintained inside the enclave. When a \lkvGet operation returns (which is potentially executed in parallel with a \lsmmerge), the enclave verifies its result against the latest file hash (updated by the \lsmmerge) in the enclave. We use an in-enclave mutex to guard the file hash under concurrent hash updates (from file replacement in a \lsmmerge) and hash reads (for verifying a \lkvGet operation).

\subsubsection{Implementation as a RocksDB Add-on}
\label{sec:impl:rocksdb}

The \elsmtwo implementation on LevelDB requires the change of LevelDB's codebase. A more modular approach (with benefits in easy maintenance and deployment, etc.) is to implement \elsmtwo without the code change of underlying LSM stores. For this reason, we present the second implementation on RocksDB. 

Comparing with LevelDB, the system of RocksDB exposes callbacks through which application programs can listen to and handle RocksDB's internal events. Different RocksDB events exposed occur at the granularity of level, file, record, etc.
The callback API is similar to stored procedures (supported in commercial database systems) and is widely available in other LSM stores (e.g., HBase Coprocessor~\cite{me:hbasecoprocessor}).

The RocksDB-based \elsm is implemented as a series of event handlers. 1) As previously described, the authenticated \lsmmerge is implemented as handlers for events \texttt{Filter} and \texttt{OnTableFileCreated}. The former event is triggered every time the underlying RocksDB encounters a key-value record during a \lsmmerge, and the latter is when a new file is written to the disk.
2) The embedded proof on RocksDB is extended to cover not only the Merkle proof at the current file, but also all Merkle proofs at the current level and previous non-hit levels. The Merkle proofs at non-hit levels are for the non-membership. 3) To implement the authenticated flush, it wraps the code for digesting a MemTable in the iterator (i.e., \texttt{next()}) exposed by a pluggable MemTable~\cite{me:rocksdbhooks}.

\ifdefined\TTUT
\section{Appendix}

\subsubsection{Multi-threading2}

\label{sec:multithread} 
Reads and writes are submitted and processed in the storage system concurrently. We assume there exists a component in the LSM store for concurrency control. For instance, LevelDB enforces single-write-multiple-reads semantics by serializing concurrent writes. This is realized by two synchronization points for write-write serialization (by a FIFO queue) and read-write synchronization (by maintaining a global timestamp). Both points are guarded by mutex. 

We place the synchronization data structures in the enclave and use the concurrency support provided in SGX SDK. On the write path, the counter update inside enclave occurs after the data write to $\level_0$ outside the enclave. On the read path, the counter read inside enclave occurs before the data read at any level outside enclave. This ensures the invariant that any execution trace will follow the order of data write, counter update, counter read and data read, which is essential to establish the freshness of data reads.
\fi

\subsection{Discussion}
\label{sec:discussion}
\subsubsection{Freshness and Rollback Attacks}
\label{sec:rollback}
In a rollback attack, the untrusted host can replace the authenticated data storage with an older but also authenticated version. To detect and defend the rollback attack, we can harden the security of \elsm systems (both \elsmone and \elsmtwo) with a trusted monotonic counter. A trusted monotonic counter provides the state freshness across power cycles. In reality, one can build a trusted monotonic counter on TPM chips (e.g., the SDK \texttt{sgx\_create\_monotonic\_counter} service~\cite{me:sdkmcounter} or others~\cite{DBLP:conf/uss/StrackxP16,DBLP:conf/uss/MateticAKDSGJC17} based on Intel Management Engine (ME)~\cite{me:intelme}) or multiple remote enclaves (e.g., as in ROTE~\cite{DBLP:conf/uss/MateticAKDSGJC17}). Given a trusted monotonic counter, the \elsm would periodically write the hash of current dataset across all levels to the counter. This hash can be implemented by the static hash of all non-zero levels (updated only when it merges) and the hash of WAL file. By this means, one can guarantee the freshness of entire dataset across power cycles against rollback attacks. The dataset-wide freshness can derive the query freshness. 

To reduce the performance impact, one can allocate a write buffer for logging the monotonic counter. The size of the write buffer is tunable by system administrator.

\subsubsection{Data Confidentiality}
\label{sec:confidential}

\elsm can be extended to support data confidentiality for applications of the need (e.g., outsourcing sensitive data to the cloud). This is handled by building an encryption layer on \elsm. While the values can be encrypted using the standard semantically-secure schemes (e.g., AES), the data keys need to be encrypted in a searchable fashion. We follow the common approach of using searchable encryption schemes~\cite{DBLP:conf/ccs/CurtmolaGKO06}, which is adopted in CryptDB and other searchable databases~\cite{DBLP:conf/sosp/PopaRZB11,DBLP:journals/pvldb/TuKMZ13}. In our design, the key of a data record is deterministic-encrypted (DE)~\cite{DBLP:conf/crypto/BellareBO07,DBLP:books/crc/KatzLindell2007}.
The use of DE allows for querying the ciphertext in the untrusted world. In our system implementation, we used the Intel SDK function \texttt{SGX\_RIJNDAEL128GCM\_ENCRYPT}~\cite{me:sgxsdk}.
For range query, one can use Order-Preserving Encryption (OPE)~\cite{DBLP:conf/eurocrypt/BoldyrevaCLO09,DBLP:conf/sp/PopaLZ13} to encrypt the data keys.
In existing work, data confidentiality in encrypted key-value stores is treated in an ad-hoc fashion (e.g., Speicher~\cite{DBLP:conf/fast/BailleuTBFHV19} is an encrypted LSM store where the correlation of block index and query is disclosed) or by more formal security protocols, e.g., Oblivious RAM, at a higher cost~\cite{DBLP:conf/ndss/AhmadKSL18}.

\subsection{Case Studies}
\label{sec:ct-sgx}

Our \elsm can be integrated with existing systems for general transparency and cryptocurrency. Here we focus on describing the case of integrating \elsm with Google's Certificate Transparency (CT) system~\cite{me:ct3}. 

A vanilla CT protocol involves with three actors, log servers, log auditors and log monitors. Log servers collect newly issued certificates (from the certificates authorities or CAs). A log auditor running along with a web browser needs to validate the certificate being used by the browser. Given a certificate, the log auditor queries the log server for a proof of inclusion (or membership) of the certificate in the CT log.
A log monitor run by a domain owner needs to monitor all certificates and to detect any certificates misissued under her domain name. 
A log monitor continuously send queries to the log server and downloads all certificates. 

With \elsm, one can build a trustworthy and efficient CT log server, which provides query authenticity and fork prevention, yet without the overhead incurred in the vanilla CT log server. More specifically, it eliminates the cost of running multiple log servers and does not need to gossip. The \elsm scheme can enable lightweight log monitors who only download the certificates of their own domain names, resulting low and sublinear bandwidth. 
In addition, with \elsm and a secure enclave, one can safely delegate the log-monitoring logic to the enclave, which relieves low-power clients (like web browser running on mobile phones) from the cumbersome auditing work. 

We have built a prototype of \elsm-based log server. In our implementation, we download certificates from the Google CT log server\footnote{\url{https://ct.googleapis.com/pilot}} and store them in \elsm by key-value records: The hostname of a certificate is used as the data key and the certificate itself (more specifically, the hash of the certificate) is the data value. \elsm receives \lkvGet queries from a log auditor and continuously monitors the certificate of interest (e.g., of a particular hostname). \elsm provides query authenticity and freshness which is essential for trusted log auditing. 

\ifdefined\TTUT
\section{Appendix}

\subsection{Alternative Designs}
\label{sec:alternatives}

\subsubsection{Place Code Outside Enclave?} 
\label{sec:codeplacement}

Both \elsmone and \elsmtwo place the entire user-space codebase of an LSM store inside enclave (note that they only differ in read buffer placement). One alternative design is to place the codebase of LSM store outside (or at least partially outside) the enclave. To be specific, the LSM store runs outside the enclave and the trusted application in enclave invokes individual \lkvPut/\lkvGet operations through OCalls.

This code-outside-enclave design requires placing ``dynamically'' updated memory data (e.g., MemTable at Level $\level_0$) outside enclave as well. Because the code outside enclave cannot access data inside enclave. In principle, one can build an authenticated data structure such as a Merkle tree over the data. However, maintaining a dynamic Merkle tree in memory incurs design complexity and sophistication especially when the data is concurrently accessed in multiple threads. We leave this class of design choices to the future work.

\subsubsection{Place Write Buffer Outside Enclave?} 
On the write path of an LSM store, the CPU stores a data update in a memory area (write buffer). Upon the buffer overflow, all data updates in the buffer are sorted in a batch before being flushed to the disk. 
Unlike the read path, the write path in an LSM store starts with data from CPU and the write buffer causes sequential disk access (instead of random page access). This characteristic makes the placement of write buffer inside enclave favorable: First, the write buffer does not benefit from being GB large\footnote{A write buffer of several data pages (4KB) is sufficiently large as it can buffer data and cause sequential disk writes.} thus placing it inside enclave will not cause expensive enclave paging. Also, securing an in-enclave write buffer is easier and more efficient than that of a buffer outside enclave, as the former can piggyback the hardware-level SGX memory protection. 
To understand the performance difference better, we present a measurement result in 
Technical Report~\cite{me:appendix}.
\fi

\ifdefined\TTUT
\subsection{Design Generalizability}
As a note for {\bf design generalizability}, we believe our work on \elsm-compatible partitioning can be applied to a broader class of systems beyond LSM-based storage systems: First, the design of \elsm-compatible partitioning is applicable to 
other emerging data storage systems designed based on data structures with minimal random writes (e.g., B$^\epsilon$ trees~\cite{DBLP:conf/spaa/BenderFFFKN07}). 
Second, we believe the \elsm design that exposes data-reorganization operations to an enclave may be applicable to various memory-management systems with a garbage-collection process, such as Java runtime, log-structured RAM storage~\cite{DBLP:conf/fast/RumbleKO14}.
Third, in principle, our design methodology in \elsm-compatible partitioning can be adopted for porting a wide range of systems to the SGX architecture with secure partitioning. 
With the \elsm partitioning, one can design a domain-specific security protocol at an appropriate abstraction level to co-locate the storage of data and that of digest/proof. Then she can partition the code and data with SGX and explore the partitioning space for better performance.
Unlike Haven style methods~\cite{DBLP:journals/tocs/BaumannPH15}, the \elsm partitioning method keeps systems-level software fully outside the enclave and significantly reduces the TCB.
\fi

\section{Performance Evaluation}
\label{sec:eva}
In this section, we present the performance evaluation of \elsmtwo and \elsmone using YCSB benchmarking tools~\cite{DBLP:conf/cloud/CooperSTRS10}.

\subsection{Experiment Design and Setup}
\label{sec:eva:setup}

{\bf Setup}: In our experiments, we use a laptop equipped with an SGX CPU. Specifically, the hardware specs include an Intel 8-core i7-6820HK CPU of 2.70 GHz with $8$ MB cache, a $16$ GB RAM and $1$ TB disk. 

We run our experiments in the YCSB framework~\cite{DBLP:conf/cloud/CooperSTRS10}. 
We use YCSB to both generate the workload and to execute the experiments. YCSB framework works in two phases: the load phase when it initializes the system by populating the dataset, and the evaluation phase when it drives the target workload to the system and measures the performance.
When initializing each experiment, we typically scan the loaded dataset so that it is loaded in the untrusted memory.
By this means, we mainly consider the setting of memory-resident data with size ranging from several hundreds of megabytes to four gigabytes (that is, millions of records, each with a 16-byte key and 100-byte value by default). With such small record size, four gigabytes is the maximal data size that we can tolerate in experiment time.

We port the open-source LevelDB-YCSB adapter~\cite{me:ycsbleveldb} to the SGX architecture. This is done by running the YCSB platform in the untrusted world and wrap each \lkvPut/\lkvGet request as an ECall (as in SGX SDK) into the enclave. Inside the enclave, we run the YCSB measurement code that measures various performance metrics.

{\bf Baseline: Eleos}: We implement a baseline of an in-memory data store. In this in-memory store, the entire dataset is stored in enclave as a sorted array. To make data update efficient, we leave $30\%$ of the array space empty to accomodate data insertions without moving existing data. For implementation, we use Eleos~\cite{DBLP:conf/eurosys/OrenbachLMS17}, a state-of-the-art virtual memory management engine in enclave without calling expensive enclave paging. Their approach, briefly, is to monitor all memory references and to relocate data between enclave and untrusted memory. We implement a sorted array in enclave linked with Eleos. The array serves data reads with binary search and is updated ``in place''. For fair comparison, the data in Eleos is persisted to disk periodically. This is done by maintaining a write buffer storing recent data updates and switching out enclave (through an OCall) for data persistence on disk.

\subsection{Overall Performance under YCSB (Macro-benchmark)}

\begin{figure*}[!htb]
  \centering
  \subfloat[Varying read-write ratio]{%
    \includegraphics[width=0.3\textwidth]{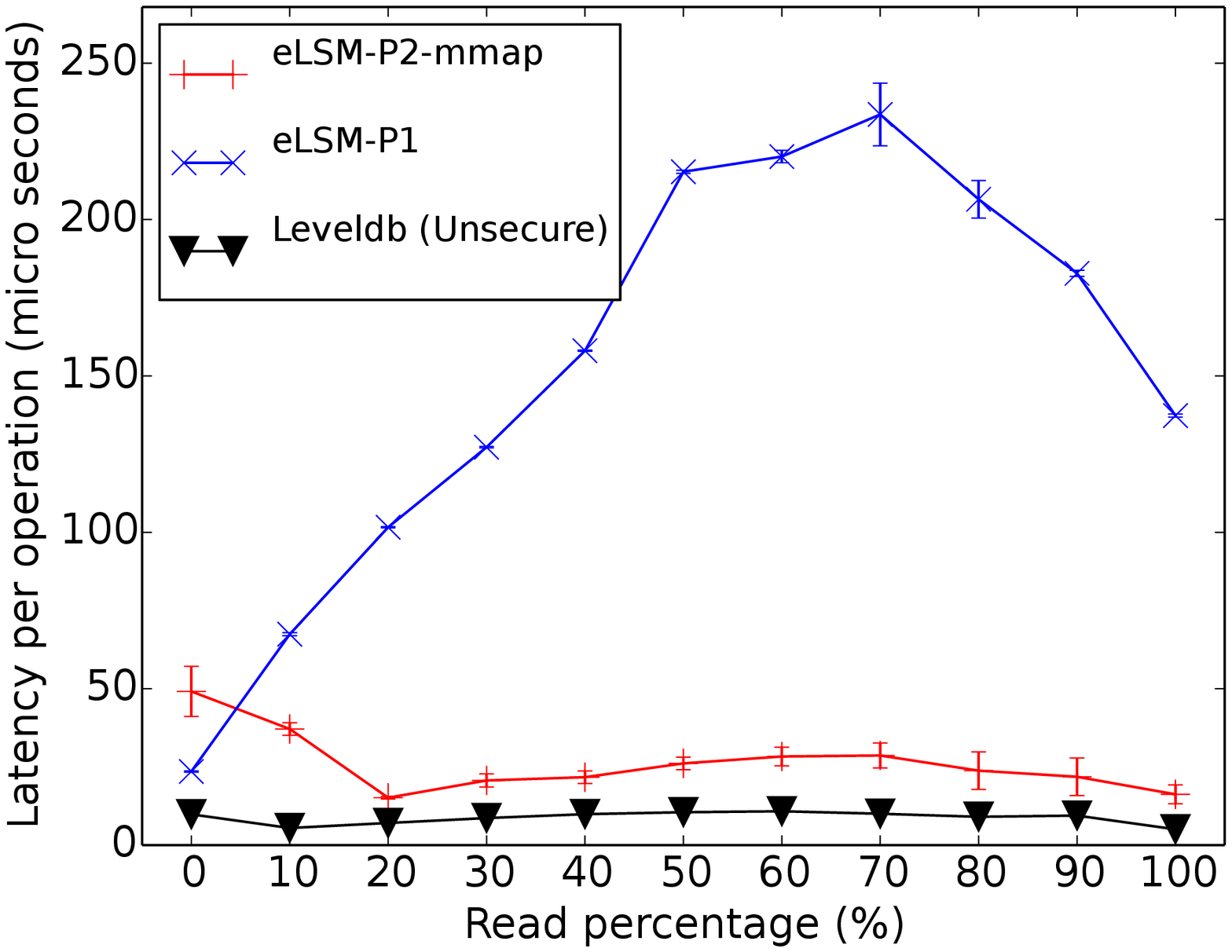}
    \label{exp:overall_rw_ycsb_ratio_preload:ss1}
  }%
  \vspace{0.1in}
  \subfloat[Varying data size (workload A)]{%
    \includegraphics[width=0.3\textwidth]{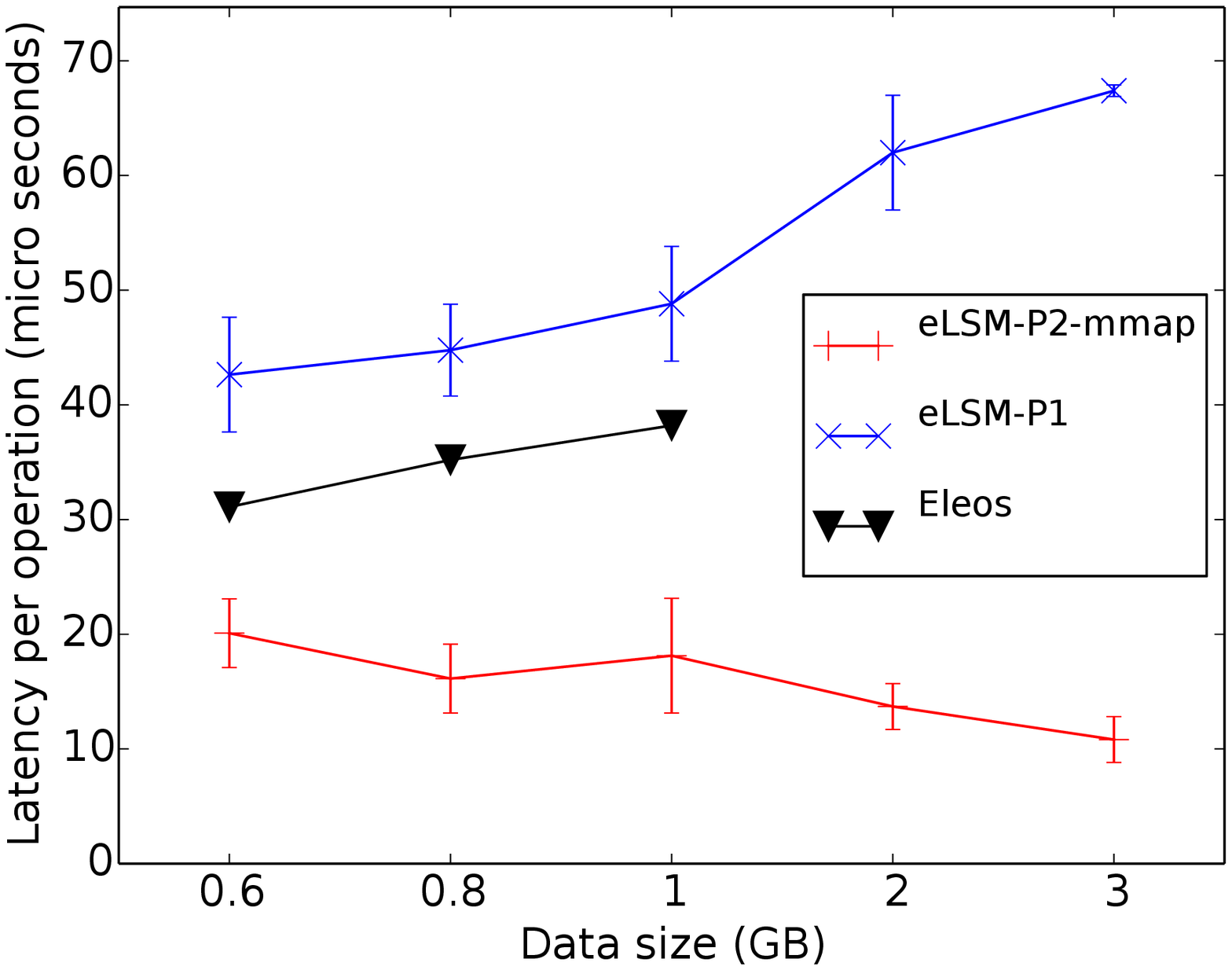}%
    \label{exp:overall_ycsb_workloada_zipfian}
  }%
  \vspace{0.1in}
  \subfloat[Varying data distributions]{%
    \includegraphics[width=0.31\textwidth]{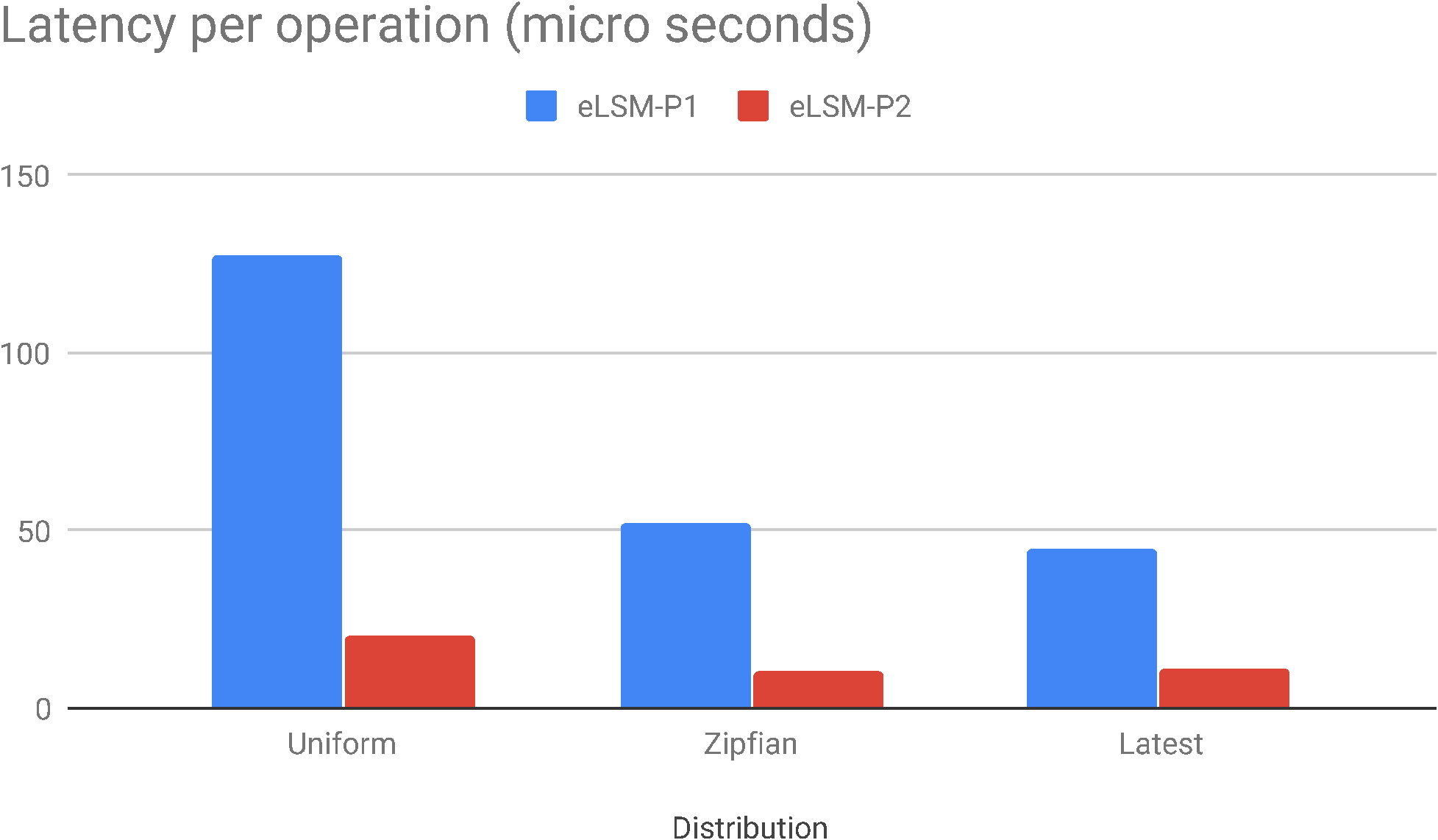}%
    \label{exp:overall_ycsb_distribution_subset}
  }%
  \caption{Performance of \elsmtwo and \elsmone under YCSB workloads}
\end{figure*}

We present the performance result of \elsm under YCSB. In this set of experiments, we vary the workloads in terms read-write ratio, key distribution, etc. and evaluate \elsm performance. The purpose is to present a holistic view regarding the performance of \elsmone and \elsmtwo.

To conduct the experiments, we fix the initial dataset at $3$ GB.
In the evaluation phase, we drive millions of operations to the key-value store for performance measurement.
We use the uniform distribution to generate the dataset and queries. 
We turn on the default compaction strategy. 
By this means, we conduct a series of experiments with varying the read-write ratio of the workload. 
Each experiment is run for three times and the average performance metric and standard deviation are reported.
The operation latency of \elsmone and \elsmtwo under varying read-write ratios are shown in Figure~\ref{exp:overall_rw_ycsb_ratio_preload:ss1}.
The result shows that \elsmtwo outperforms \elsmone in most workloads except for a small set of write-only workloads. Specifically, as the workload becomes more read intensive, \elsmtwo (with the \texttt{mmap} configuration as described in \S~\ref{sec:mmap}) has its operation latency decreased. This performance characteristic is due to that \elsmtwo has to cause disk IO for data persistence on the write path while on the read path it can read the memory (through the \texttt{mmap} files). As the workload transitions from writes to reads, \elsmone's latency first increases and then decreases near the end. The increase of latency is caused by overflowing the enclave memory (of 128 MB) and enclave paging, as will be validated in other experiments (e.g., in Figure~\ref{exp:read:data_size:ss1}). In addition, compared with the ideal approach (running an unsecured LevelDB), the slowdown caused by \elsmtwo ranges between $1.5\times$ and $4X$. 

Comparing \elsmtwo and \elsmone, when the workload is write-only, \elsmone is faster. 
For most workloads, \elsmtwo has a smaller operation latency than \elsmone, and the performance discrepancy reaches the highest when the workload consists of $70\%$ reads (Note the uniform key distribution in this workload). In this setting, \elsmtwo achieves $4.5X$ performance speedup comparing \elsmone.
This performance result clearly supports the design tradeoff made in \elsmone and \elsmtwo, where \elsmtwo optimizes the read path by placing the read buffer outside enclave and avoiding enclave paging, which inevitably causes the write overhead, including authenticating \lsmmerge and embedding \elsmtwo proofs in the software layer. \elsmone does not have such write overhead (data security is provided by the hardware-level memory protection in SGX). From the performance result, it can be seen that the \elsmtwo's design to trade off write performance for read is worthwhile, as the majority of workloads favors \elsmtwo. 

The second experiment is to report the operation latency under varying data sizes. 
We initialize the system with data of varying sizes from $0.6$ GB to $3$ GB. 
In the evaluation phase, we drive into the system YCSB workload A which consists of $50\%$ reads and $50\%$ writes with data keys generated following a Zipfian distribution. We measure the operation latency for \elsmtwo (in \texttt{mmap} configuration), \elsmone and the baseline of Eleos. The result is shown in Figure~\ref{exp:overall_ycsb_workloada_zipfian}. 
With the increasing data sizes, 
Eleos can scale only to $1$ GB data which is limited by their open-source project~\cite{me:eleoscode,DBLP:conf/eurosys/OrenbachLMS17}.
The discrepancy between the latency of \elsmtwo and \elsmone increases, which reaches a maximal of $7X$ difference when the data size is $3$ GB. 

Figure~\ref{exp:overall_ycsb_distribution_subset} presents the operation latency when the workload is generated with different key distributions. YCSB provides three common distribution for generating data keys, that is, Uniform, Zipfian and Latest. Among the three, Latest is the key distribution that has the best temporal locality (as it tends to read the latest inserted records), leading to smaller working sets. 
In this experiment, we use an initial dataset of $3$ GB. In general, \elsmtwo is less sensitive to key distribution than \elsmone. Under the uniform distribution, \elsmone causes the highest operation latency. Because the working set size is the largest when data keys are generated uniformly, and it causes the highest memory pressure in enclave in \elsmone's design.

\subsection{
Read Performance}

\begin{figure*}[!htb]
  \centering
  \subfloat[Memory placement under varying data sizes]{
    \includegraphics[width=0.30\textwidth]{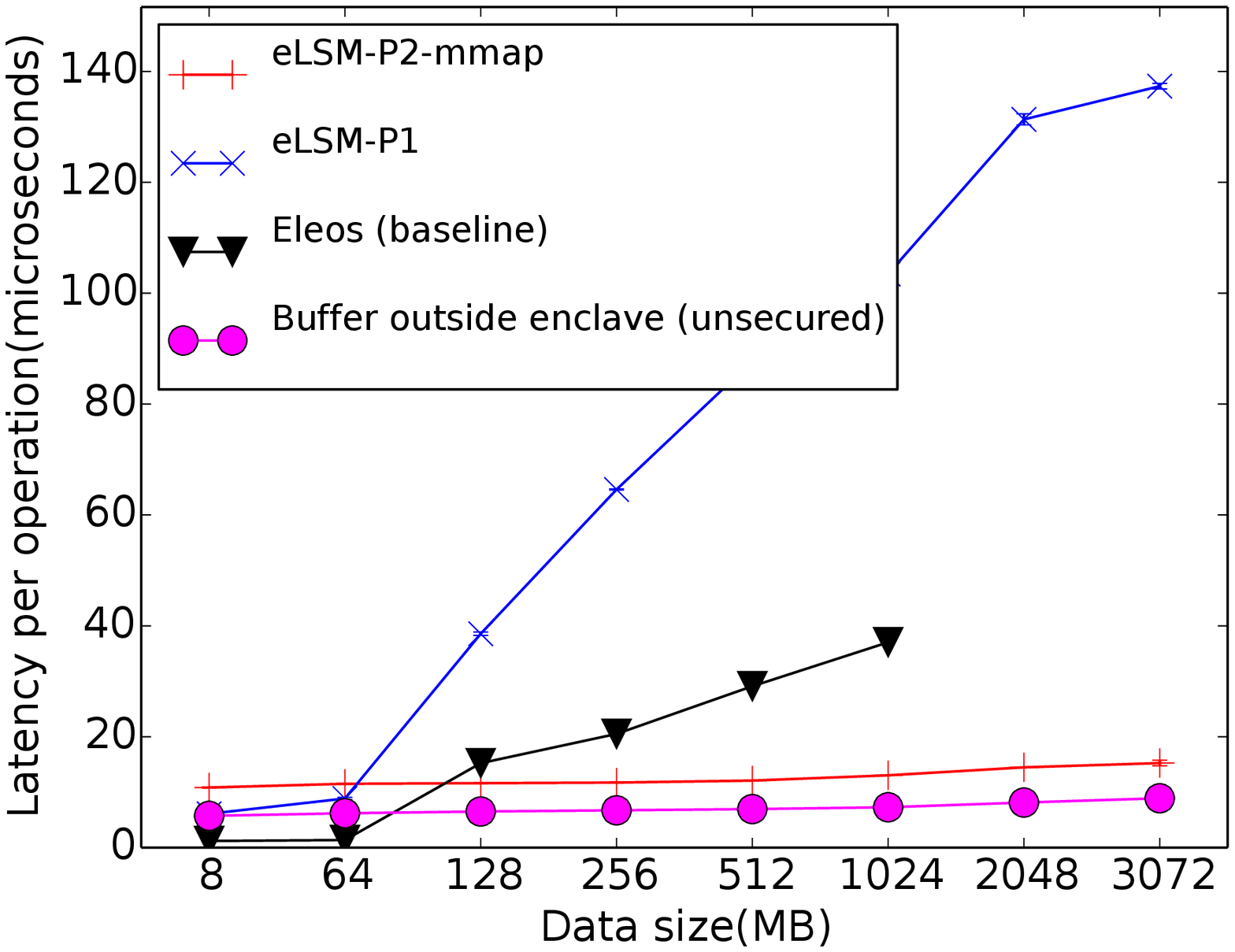}
    \label{exp:read:data_size:ss1}
  }
  \vspace{0.1in}
  \subfloat[Comparing \texttt{mmap} and buffer in \elsmtwo]{%
    \includegraphics[width=0.30\textwidth]{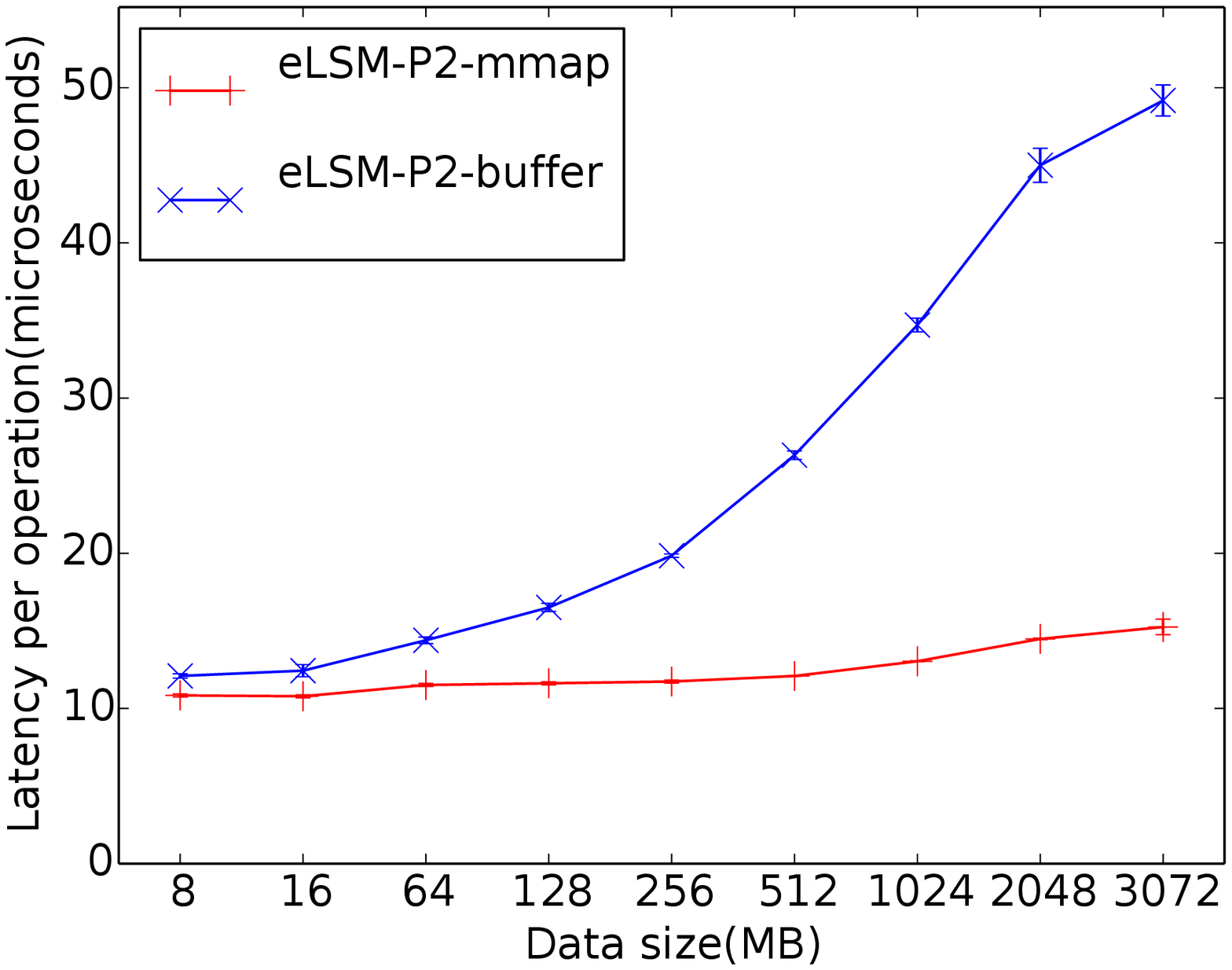}
    \label{exp:read:data_size:ss2}
  }%
  \vspace{0.1in}
  \subfloat[Varying buffer size]{%
    \includegraphics[width=0.30\textwidth]{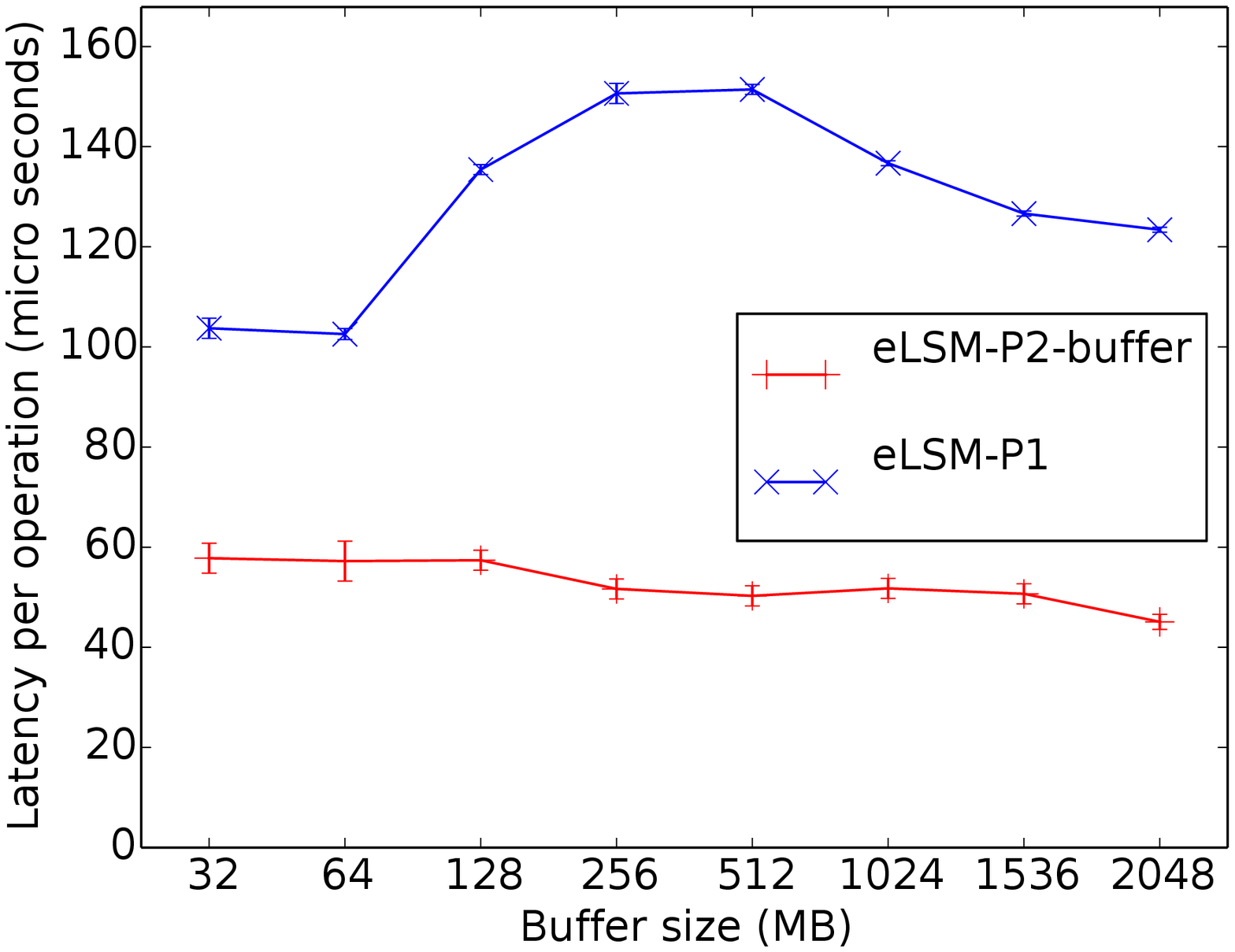}%
    \label{exp:overall_ycsb_readonly_uniform_2g_buffer_preload}
  }%
  \caption{Read performance of \elsmtwo, \elsmone and the Eleos baseline}
\end{figure*}

This set of experiments evaluate the read performance of \elsmtwo and \elsmone.
We initialize the storage system with datasets of varying sizes from $8$ MB to $3$ GB. We then drive
a read-only workload of one million \lkvGet requests and measure the latency of these requests.
In addition to Eleos, we consider another baseline that was mentioned in our initial performance study (in Figure~\ref{fig:eval:motive1}). The baseline places the read buffers outside the enclave (similar to \elsmtwo) but does not have the data-authentication measures including the \elsm Merkle trees and proofs (unlike \elsmtwo). 
This baseline is thus unsecured but it serves to show the ideal performance. 
We use the \texttt{mmap} configuration in \elsmtwo which allows for accessing files pinned in memory. 

The read latency is presented in Figure~\ref{exp:read:data_size:ss1}.
When the data size is smaller than the enclave memory size (i.e., smaller than 128 MB), \elsmone and the baseline of Eleos perform better than \elsmtwo because \elsmtwo incurs proof and verification in the software layer. When the data size grows beyond the enclave memory (i.e., larger than 128 MB), \elsmtwo outperforms \elsmone and Eleos. Again, Eleos limits the data scalability to $1$ GB data. The longer latency in Eleos may be caused by the overhead for runtime monitoring and extra data copy in enclave memory.
\elsmtwo generally keeps the read latency constant with the increasing data size.

On the read path, \elsmtwo can support both \texttt{mmap} files and a user-space buffer. Note that \elsmone cannot support \texttt{mmap} files as such files must reside in the kernel-space memory outside the enclave.
We compare the performance of \texttt{mmap} read and buffer-based read in \elsmtwo. The result is in Figure~\ref{exp:read:data_size:ss2}. As data grows, the performance advantages of \texttt{mmap} configuration in \elsmtwo become clearer. At the largest data scale tested, \elsmtwo with \texttt{mmap} achieves $5X$ speedup of read latency comparing with the read buffer configuration.

We also compare \elsmtwo (the buffer configuration) with \elsmone under varying buffer sizes. Both \elsmtwo (buffer) and \elsmone support user-space memory buffers and this experiment is intended for a fairer comparison. In the experiment, we fix the data size to be $2$ GB and vary the buffer size from $32$ MB to $2$ GB. 
The result is in Figure~\ref{exp:overall_ycsb_readonly_uniform_2g_buffer_preload}.
It can be seen that \elsmtwo stays constant with increasing buffer size (Note that the data size is fixed and this is a different setting from the previous experiment in Figure~\ref{exp:read:data_size:ss2}) and \elsmone's read latency increases sharply around 128 MB data size. In general, \elsmtwo (buffer) achieves $1.6X\sim{}2.3X$ speedups against \elsmone.

\subsection{
Write Performance}
\label{sec:putonly}

\begin{figure}[!htb]
  \subfloat[Comparing \elsmtwo, \elsmone and Eleos]{%
    \includegraphics[width=0.25\textwidth]{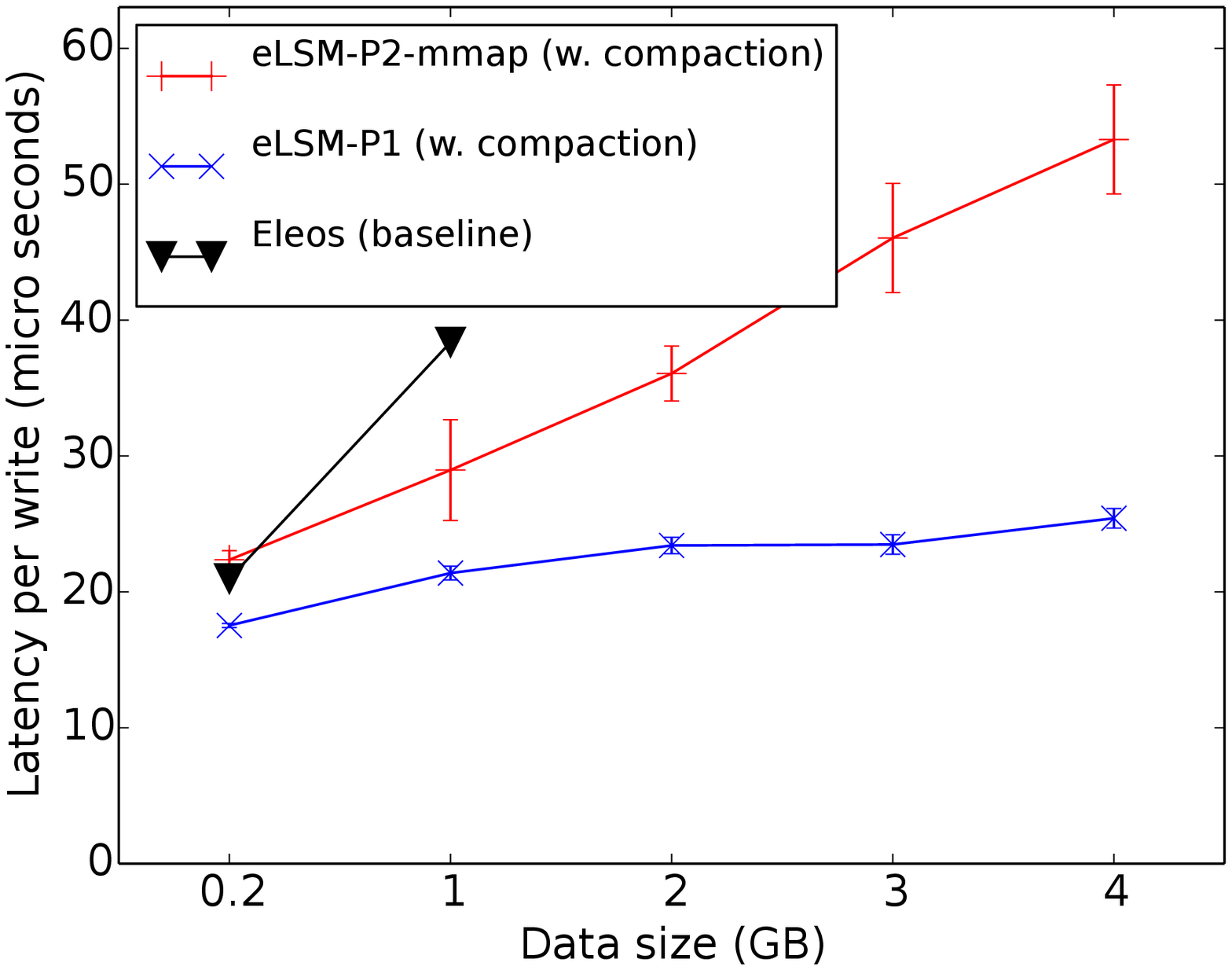}%
    \label{exp:overall_writeonly_new:2}
  }%
  \subfloat[Writes w/wo compaction]{%
    \includegraphics[width=0.25\textwidth]{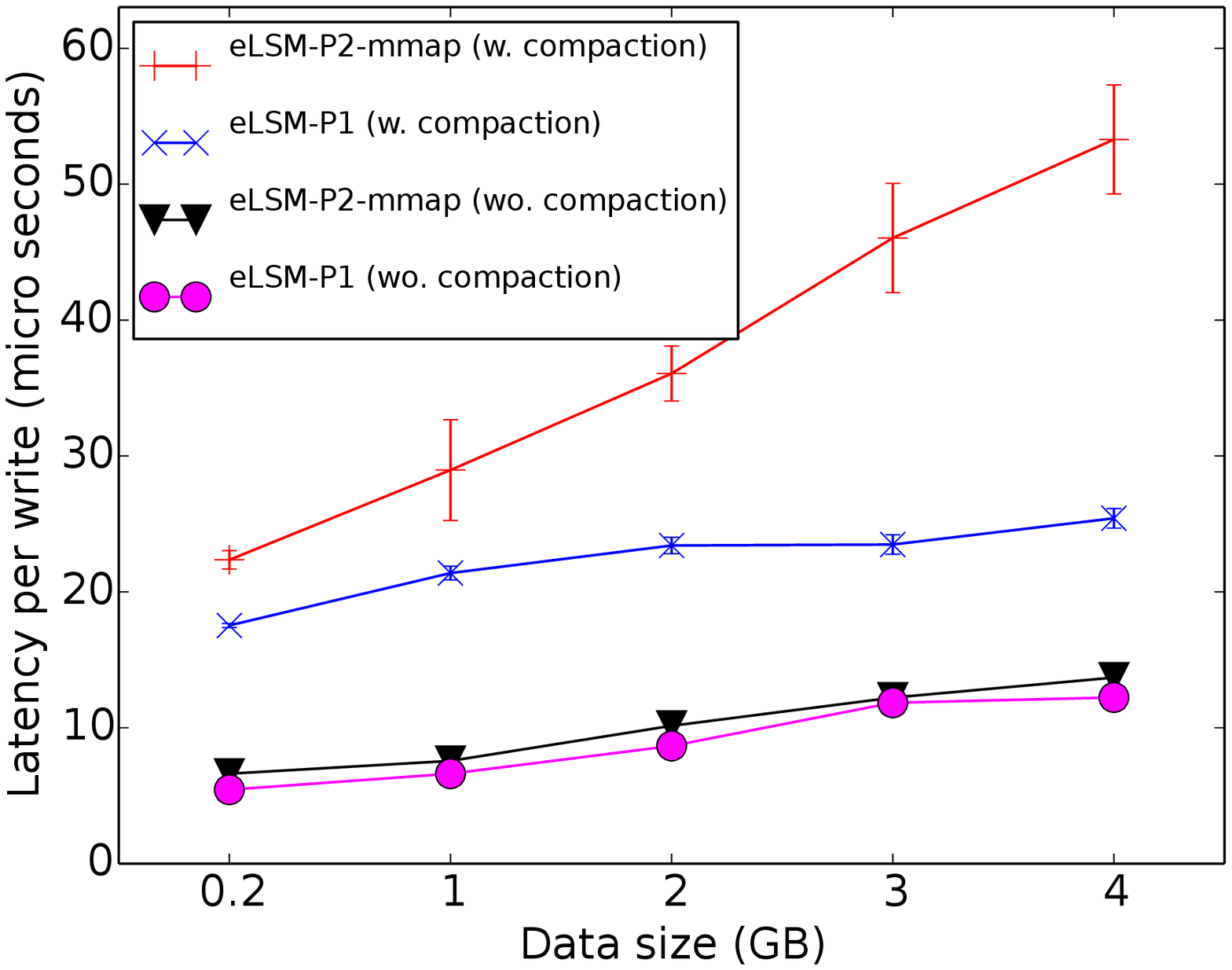}%
    \label{exp:overall_writeonly_new:1}
  }%
  \caption{Write performance with/without \lsmmerge in \elsmtwo and \elsmone}
\end{figure}

This set of experiments measure the write performance of \elsmtwo and \elsmone. In the experiments, we initialize the system with dataset of varying sizes from $0.2$ GB to $4$ GB. In the evaluation phase, we measure the write latency. For \elsmtwo and \elsmone, we consider the configurations with and without \lsmmerge. We also consider the baseline of Eleos.

The write performance with \lsmmerge is reported in Figure~\ref{exp:overall_writeonly_new:2}. We report as the write latency the average execution time of running an \lkvPut request plus the time for \lsmmerge amortized to the individual \lkvPut. Among the three approaches, \elsmone is the fastest on the write path. With \lsmmerge turned on, \elsmtwo's write latency is about $1.3X\sim{}2.3X$ of that in \elsmone. Eleos is the slowest and can only scale to the data size of $1$ GB. Because in both \elsmtwo and \elsmone, the memory footprint in the write path is limited to a small write buffer (of $4$ MB), whereas the Eleos is an update-in-place structure with the working set size equal to the data size. Even without hardware-level enclave paging (as is optimized out by Eleos), the update-in-place structure still incurs a large amount of memory copies across the enclave boundary, leading to a higher write latency than \elsmtwo and \elsmone.

We also compare the write performance with and without \lsmmerge. The result is presented in Figure~\ref{exp:overall_writeonly_new:1}. It can be seen that turning on \lsmmerge, it causes $2X\sim{}4X$ slowdown on the write path. In both cases, namely with and without \lsmmerge, \elsmtwo has a higher write latency than \elsmone. This is due to the overhead caused by building the embedded proof in \elsmtwo.

\ifdefined\TTUT

\subsubsection{
\lsmmerge Performance}

\begin{figure}[!htb]
\centering
    \includegraphics[width=0.3\textwidth]{figures_ju_socc19/compaction-subset1.eps}
  \caption{\lsmmerge performance}
  \label{exp:compaction2}
\end{figure}

\subsection{Multi-threading in YCSB}

\begin{figure*}[!htb]
  \centering
  \subfloat[Read latency]{%
    \includegraphics[width=0.25\textwidth]{figures_ju_socc19/latency_mem_read.ps}%
    \label{exp:latency_mem_read}
  }%
  \subfloat[Read throughput]{%
    \includegraphics[width=0.25\textwidth]{figures_ju_socc19/throughput_mem_read.ps}%
    \label{exp:throughput_mem_read}
  }%
  \subfloat[Write latency]{%
    \includegraphics[width=0.25\textwidth]{figures_ju_socc19/latency_mem_write.ps}%
    \label{exp:latency_mem_write}
  }%
  \subfloat[Write throughput]{%
    \includegraphics[width=0.25\textwidth]{figures_ju_socc19/throughput_mem_write.ps}%
    \label{exp:throughput_mem_write}
  }%
  \caption{Performance with varying numbers of threads (YCSB - multi-threads, memory data (1GB); default compaction; system exp.); NEW}
\end{figure*}

\fi

\section{Related Work}
\label{sec:rw}


{\bf Software Systems on SGX}: Since the release of Intel SGX, there has been a large body of research on building software systems on SGX. The existing works have addressed the in-enclave support for OS (e.g., Haven~\cite{DBLP:conf/osdi/BaumannPH14}, Graphene-SGX~\cite{DBLP:conf/usenix/TsaiPV17}, SCONE~\cite{DBLP:conf/osdi/ArnautovTGKMPLM16}, and Panoply~\cite{shinde2017panoply}), security applications~\cite{DBLP:conf/usenix/LindPMOAKRGEKFP17}, side-channel attacks~\cite{DBLP:conf/sp/XuCP15,DBLP:conf/uss/BulckWKPS17,DBLP:conf/usenix/HahnelCP17,DBLP:journals/corr/BrasserMDKCS17,DBLP:conf/uss/0001SGKKP17} and defenses~\cite{shih2017t,DBLP:conf/ccs/ShindeCNS16,DBLP:conf/ccs/ChenZRZ17}, databases~\cite{DBLP:conf/sp/SchusterCFGPMR15,DBLP:conf/uss/OhrimenkoSFMNVC16,enclavedb-a-secure-database-using-sgx,DBLP:conf/nsdi/ZhengDBPGS17,DBLP:journals/pvldb/BajajS13}, etc. 

In particular, there is a line of research on building key-value stores in enclave. Pesos~\cite{DBLP:conf/eurosys/KrahnTVKBF18} supports the secure hosting of a key-value store where the enclave enforces fine-grained access-control policies.
SecureKeeper~\cite{DBLP:conf/middleware/BrennerWGWLFPK16} secures ZooKeeper style coordination services hosted in the public cloud. Their approach is to confine the computation of user-provided data in enclave and to encrypt data in transit. 
Eleos~\cite{DBLP:conf/eurosys/OrenbachLMS17} supports in-memory key-value stores and particularly addresses the problem of fitting in enclave data larger than 128 MB. They provide a virtual-memory abstraction in enclave and optimize out the expensive enclave paging by monitoring user-space memory accesses and by relocating data dynamically between the enclave memory and untrusted memory. 
HardIDX~\cite{DBLP:conf/dbsec/FuhryBB0KS17} is a secure index in enclave that seals external data using authenticated encryption~\cite{DBLP:books/crc/KatzLindell2007}.
Concerto~\cite{DBLP:conf/sigmod/ArasuEKKMPR17} supports concurrent key-value stores with consistency guarantees. It is based on a novel design to check strong consistency by leveraging homomorphic secure hash with enclave.
Shieldstore~\cite{DBLP:conf/eurosys/KimPWJH19} supports in-memory key-value stores by placing data outside enclave and by running inside enclave an engine for record-grained encryption and integrity-checking. Shieldstore does not support data persistence or LSM tree. 

{\bf LSM Storage Systems}: 
bLSM~\cite{DBLP:conf/sigmod/SearsR12} optimizes the LSM tree performance by row-based data storage and fine-grained compaction.
Prior work~\cite{DBLP:journals/vldb/JermaineOY07} minimizes the write amplification under the skewed key access pattern. 
Pebble~\cite{DBLP:conf/sosp/RajuKCA17} reduces the write amplification by organizing storage layout in skip lists and avoiding data rewriting in the same level. 
Accordion~\cite{DBLP:journals/pvldb/BortnikovBHKS18} applies the principle of LSM tree to the memory management in order to solve the write amplification caused by frequent compaction.
Compaction is critical to the performance of LSM tree storage. Existing work studies distributed compaction management in a cluster setting~\cite{DBLP:journals/pvldb/AhmadK15}.
Beyond disk storage, the LSM tree has been applied for main-memory databases with high compression rate~\cite{DBLP:conf/sigmod/ZhangAPKMS16}, on non-volatile memory~\cite{DBLP:conf/usenix/MarmolSTR15}, and for spatial databases in the AsterixDB project~\cite{DBLP:journals/pvldb/AlsubaieeBBHKCDL14}. Concurrency of the LSM tree is studied in cLSM~\cite{DBLP:conf/eurosys/Golan-GuetaBHK15} that supports snapshot scan, conditional update, and concurrent \lsmmerge.

\label{sec:vsspeicher}

{\bf Comparison with Speicher}: Speicher~\cite{DBLP:conf/fast/BailleuTBFHV19}, published recently, is a secure system of LSM store with Intel SGX. It places inside the enclave the data keys in MemTable and meta data in other levels to enable search. Data values are stored outside the enclave. Speicher presents a series of performance optimization for efficient IO in SGX.

Our \elsm is developed independently with Speicher. Comparing with it, \elsm has technical distinctions listed as following: 
1) \elsm supports efficient data reads (\lkvGet) with early stops. As mentioned in \S~\ref{sec:readwrite}, \elsm stops processing a \lkvGet as early as the first hit is found, that is, without scanning all the levels. This leads to significant performance improvement especially for workloads exhibiting temporal locality (recall Figure~\ref{exp:overall_ycsb_distribution_subset}). Also, the capability is much needed in the context of incremental log monitoring where only the recent records are of interest (see the case study in \S~\ref{sec:ct-sgx}). In addition, we present non-trivial security analysis of this capability (i.e., Theorem~\ref{thm:security} and Lemma~\ref{thm:temporalorder}). By contrast, existing work including Speicher requires each \lkvGet to iterate through all levels.
2) We present a modular implementation of \elsm without changing the code of RocksDB. This is realized by the \elsm design in embedding proofs in data records and leveraging the callback hooks in RocksDB (\S~\ref{sec:impl:rocksdb}). 
By contrast, existing work including Speicher requires code modification of the underlying LSM stores~\cite{DBLP:conf/fast/BailleuTBFHV19}.
3) We have integrated \elsm in Google's certificate transparency for a real-world case study (\S~\ref{sec:ct-sgx}). We also open-source the code of \elsm~\cite{me:elsmcode}.

At last, Speicher supports encrypted data keys in the untrusted world. To search the encrypted LSM store, it needs to ``correlate the block'' index of an SSTable with the queried key. As the block index being accessed is leaked to the untrusted world, its correlation to data key may lead to information leakage and break the query confidentiality. This security implication of this leakage however is not well documented in the paper.

\section{Conclusion}
\label{sec:conclude}

This paper presents a novel SGX-based LSM key-value store to address emerging application needs of authenticating data under frequent updates.
The proposed \elsmtwo system places memory buffers outside enclave for efficiency. It authenticates the data with small proofs at selective levels of an LSM tree. Implementation on LevelDB and RocksDB is presented. The performance studies with YCSB workloads show a $4.5X$ speedup of \elsmtwo over the baseline of \elsmone.

{
}

\newpage


\bibliographystyle{abbrv}

{
\bibliography{ads2,crypto,distrkvs,lsm,sc,txtbk,yuzhetang,bkc,diffpriv,latex,odb,sgx,t_tee,vc}
}

\appendix
\section{Additional Preliminaries}
\label{sec:prel:sgx}
\subsection{Software Guard eXtension (SGX)}

Intel SGX is a security-oriented x86-64 ISA extension on the Intel Skylake CPU, released in 2015.
SGX provides a ``security-isolated world'' for trustworthy program execution on an otherwise untrusted hardware platform.
At the hardware level, the SGX's trusted world or enclave includes a tamper-proof CPU which automatically encrypts memory accesses upon cache write-backs. Programs executed outside the enclave trying to access enclave memory only get to see the ciphertext and cannot succeed. 
At the software level, the SGX enclave includes only some unprivileged program and excludes any OS kernel code, by explicitly prohibiting system services (e.g., system calls) inside the enclave.

To use the technology, a client initializes an enclave by uploading her program to the server host and uses SGX's seal and attestation mechanism~\cite{Anati_innovativetechnology} to verify the correct setup of the enclave environment (i.e., the binding between the client's program and a genuine CPU supporting SGX). During the program execution, the enclave can be entered and exited proactively (by SGX instructions, e.g., \texttt{EENTER} and \texttt{EEXIT}) or passively (by interrupts or traps). These world-switch events trigger the context saving/reloading in both hardware and software levels.
Comparing prior TEE solutions~\cite{me:txt, me:tpm, me:tzone, me:scpu}, SGX uniquely supports multi-core concurrent execution, dynamic paging, and interrupted execution.

The software built on SGX relies on an underlying ``RPC'' mechanism to switch the execution between the enclave and untrusted world. In SGX SDK, such RPC mechanism is supported by ECall/OCall where in the case of an ECall (OCall), the untrusted host (enclave) switches to enclave (the untrusted host) in order to call a function there. Alternatively, one can use the LibOS supports in enclave to load and run unmodified application software in enclave. There are existing research prototypes for enclave LibOS, such as Haven~\cite{DBLP:journals/tocs/BaumannPH15}, Graphene-SGX~\cite{DBLP:conf/usenix/TsaiPV17}, and SCONE~\cite{DBLP:conf/osdi/ArnautovTGKMPLM16}. 

{\bf SGX storage and costs}: In SGX, there is a reserved region of main memory 
where the CPU stores data encrypted and authenticated. This CPU-protected region of memory, called PRM, is small and in current Intel CPU, the size is limited to 128 MB. The SGX architecture supports a virtual memory in enclave of arbitrary size. When the data stored in enclave virtual memory is larger than 128 MB, it causes the event of ``enclave paging'', that is, the hardware will transfer the data between  untrusted main memory and the protected memory region (PRM). The enclave paging is expensive as the process involves asynchronous enclave exit (i.e., \texttt{AEX}), OS-managed page table, and SGX instructions (i.e., \texttt{EWB}) to evict a victim page to remap the requested page. This enclave paging event can easily become the system bottleneck when the application has a working set larger than 128 MB.

\ifdefined\TTUT 
Very recently, there are system works that optimize the support of large dataset in enclave. Notably, Eleos~\cite{DBLP:conf/eurosys/OrenbachLMS17} provides a user-managed virtual memory in enclave by monitoring the enclave memory utilization and automatically copying data across the enclave boundary. However, Eleos supports coarse-grained data protection (page level) and does not support efficient key-value stores~\cite{DBLP:conf/eurosys/KimPWJH19}.
\fi

\subsection{Merkle Hash Trees}

A Merkle Hash Tree (or MHT)~\cite{DBLP:conf/sp/Merkle80} is a method of digesting and authenticating a dataset. Specifically, given an array of data records, a Merkle tree's leaf set consists of the hashes of all data records. The non-leaf node in a Merkle tree is the hash of the concatenation of the children of this node. The root hash of the Merkle tree digests the entire data array with records in the fixed order. Merkle trees are often used in constructing a proof system between a verifier and a prover. Given a query over a dataset, a Merkle proof consists of all hashes of the tree nodes surrounding the path from the leaf (that matches the query) to the root. Hence, it is also called Merkle authentication path. When the Merkle tree is built over a sorted dataset, the Merkle proof can be used to authenticate the membership and non-membership of a query result in the dataset. In particular, when there is no matching record to the query, the Merkle proof for non-membership consists of the two authentication paths respectively for the two records immediately larger and smaller than the queried data.

\section{Detailed Target Scenarios and Workloads}
\label{sec:targetapp2}

In practice, the LSM based key-value stores are being used increasingly in security sensitive applications. As an example, LevelDB serves as the data storage solution in security-critical scenarios including certificate transparency log~\cite{me:ct,DBLP:conf/uss/MelaraBBFF15}, Blockchain/cryptocurrency~\cite{me:bitcoin,me:eth,me:litecoin}, web browser storage~\cite{me:chrome}, etc. In the following, we give a more detailed description of these scenarios before summarizing the common traits in these target applications.

{\bf In cryptocurrency and Blockchain applications}, LSM stores serve as the ledger storage. For instance, LevelDB is adopted in Bitcoin~\cite{me:ldbbitcoin}, Ethereum~\cite{me:ldbethereum}, HyperLedger~\cite{me:ldbhyperledger}, multichain/stream~\cite{me:ldbmultichain}, etc. 
These are user-space storage systems that translate application-level reads/writes (on individual data records) to lower-level file reads/writes. With an LSM tree, they do so by translating random writes to sequential file writes, while still supporting efficient processing of random-access reads. Specifically, an LSM tree represents a dataset \m by a series of levels $\level_0,\level_1,...$ where each level consists of key-value records sorted by data keys. Upon a write operation, the LSM tree only updates the first level $\level_0$ in memory. When the memory buffer at $\level_0$ is full, it flushes all buffered recent writes (sorted in order) to the disk, creating a new file in $\level_1$. Periodically, the LSM tree will merge multiple levels into one, by accessing all files/records of involved levels in batch. By this means, on the write path, it only cause sequential file access, which reduces disk seeks.

{\bf In certificate-transparency log} (CT log~\cite{me:ct}), LSM stores serve as a publicly verifiable storage of certificates~\cite{me:ldbct}. A certificate is stored in the CT log with its identity as the key and the certificate as the value. It serves the workloads of data updates by inserting or revoking certificates, and the read queries to retrieve certificates by identity. The data integrity of a certificate is critical to the security in CT, as using a revoked certificate may lead to the misuse of a stolen key and facilitating impersonation attacks.

{\bf Web browser storage} is one of the canonical applications of LSM stores. LevelDB is primarily used as the storage backend in Google Chrome, responsible for storing a large amount of structured data~\cite{me:ldbchrome}. The data stored in LevelDB in Chrome can be personal files and information, used in Browser extensions. For instance, a Chrome extension for supporting Blockchain (i.e., Nimiq~\cite{me:ldbnimiq}), LevelDB stores the Blockchain data through the web interface. In these applications, serving the fresh data with integrity is security critical (as mentioned above).

{\bf User-generated content}:
For an example, suppose a Twitter-alike startup company wants to outsource the storage of tweets to a third-party cloud (e.g., for economic reasons). The application-level web server accepting user requests will generate two types of queries for the outsourced storage underneath: 1) writes of new tweets as social users post them, 2) reads of recent tweets to analyze trendy stories and to present them in the ``Moments'' page (on a Twitter-alike website). 
This persistence workload features a stream of small writes that arrive at a high rate and reads that need to be served in a real-time fashion. An LSM tree is suited to serve this workload -- with the first level ($\level_0$) of constant size and a constant number of levels, the LSM tree can serve intensive writes with small memory and bound the latency of a read in quasi-linear time. The data authenticity in this Twitter application means that social users will neither be fooled by a fake post nor miss their friends' newest update. 

{\bf Streaming data analytics}: 
For another example, many big-data applications, such as real-time monitoring and analytics (e.g., Spark), feature an intensive volume of data reads and writes as a typical ``big-data'' attitude is to collect everything first and to process it later~\cite{Schneier:2015:DGH:2685412}. They require to minimize the time between data collection and processing for real-time analytics. An LSM tree is well suited to serve big-data storage as its log-structured design allows to ``collect everything'' at high write throughput and its merge design allows to ``process'' the data as soon as possible. 


We summarize the common traits in the above application scenarios and formulate our work in terms of target workloads (\S~\ref{sec:workloads}) and security requirements (\S~\ref{sec:security}).

\subsubsection{Target Workloads}
\label{apx:workloads}

In our target applications, the workload features intensive write streams and random-access reads. For instance, Blockchain applications feature an intensive stream of incoming cryptocurrency transactions and random-access queries to selectively download relevant blocks/transactions when new Blockchain nodes join the network (as in lightweight SPV nodes~\cite{me:spv}). CT log's read/write workloads features an intensive stream of updates, especially in a large-scale setting such as key transparency scheme (CONIKS~\cite{DBLP:conf/uss/MelaraBBFF15}). CT log serves random-access reads to selectively retrieve the certificate of interest.

In general, our target workload is characterized as below: Our data model is a key-value store, where each record contains a data key through which the record is accessed. On the read path, the workload supports read queries where each read request specifies the data key and retrieves the associated value. The read queries cause random access as the data key can be arbitrarily distributed (C1). 

On the write path, there is an \emph{intensive} stream of data writes that need to be persisted in real time. The stream consists of small random-access writes where keys are arbitrarily distributed (C1). To be specific, a data write manifests in an insertion of a key-value record, where keys can be arbitrarily distributed. The target workload features an intensive write stream, requiring the underlying storage system to provide high throughput and low latency. 

This target workload excludes many database storage engines that are designed based on update-in-place data structures (e.g., B+ trees). In update-in-place structures, a data update needs to overwrite the previous version of the record at the exact location where the record is stored. An update incurs lookups and random-accesses of the record's previous location, leading to disk seeks and write amplification. The LSM stores are suited for the target workload of intensive small writes (C2), due to its append-only design that persists the writes in their arrival order. 

\section{Write Buffer Placement: Measurement Results}
\label{apx:writebufferin}

\begin{figure}[!hbtp]
    \centering
    \includegraphics[width=0.35\textwidth]{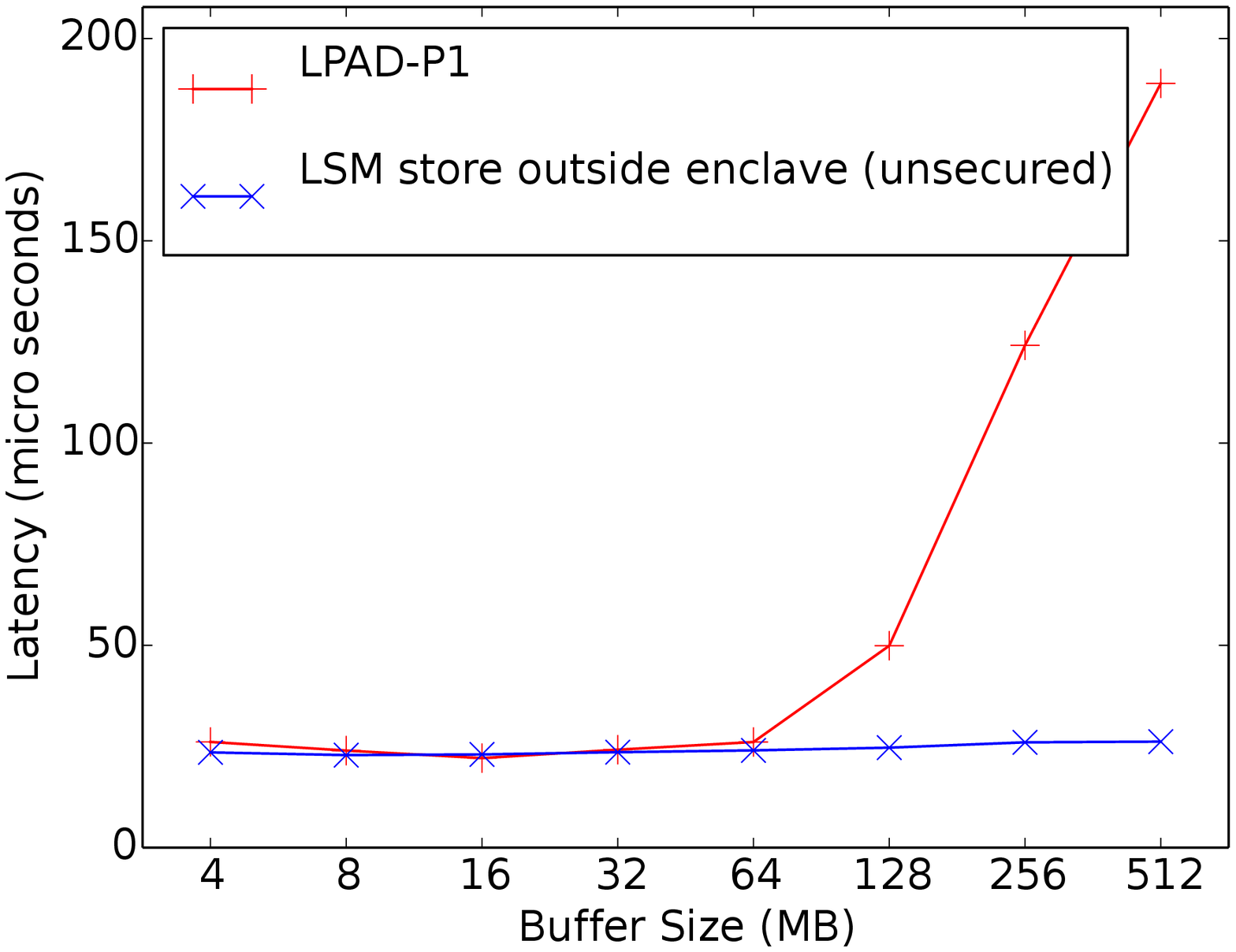}
    \label{fig:write_buffer_size}
    \caption{Disk writes: Placing data outside}
\end{figure}

We conduct a preliminary performance study of placing the write buffer inside enclave (i.e., \elsmone) in comparison with that outside enclave. The latter is implemented by the enclave issuing data writes to the untrusted memory (even without security protection) and flushing the write buffer with a world switch. In the performance result, it can be seen that 1) for an unsecured LSM store, a small write buffer results in a similar performance with a large write buffer. 2) With a small buffer, placing the buffer inside enclave and outside enclave result in the similar performance. In other words, placing the write buffer outside enclave does not improve the performance, yet incurs much higher implementation complexity. 

\section{Alternative Design: TCB-minimal Code Placement}
\label{sec:codeplacement2}

To decide which part of the codebase can be placed outside enclave, we take the following view: The codebase of a storage system consists of ``computation''-oriented code and data-access oriented code. In an LSM store, there are two major computations: the merge operation and the versioning (the latter is about selecting and comparing timestamps among different records of the same key). When placing the computation-oriented code outside enclave, it entails very expensive cryptographic protocols/schemes to ensure the computation security. For instance, to ensure the integrity of computation result, it requires running verifiable computation protocols~\cite{DBLP:journals/jacm/AroraLMSS98} in the untrusted world, whereas the state of the art systems for verifiable computation~\cite{DBLP:conf/sp/ParnoHG013,DBLP:conf/eurosys/SettyBVBPW13,DBLP:conf/sosp/BraunFRSBW13,DBLP:conf/ndss/WahbySRBW15} cause performance slowdown (comparing unsecured computation) by multiple orders of magnitudes. Because of this, we have to place the computation-oriented code inside the enclave.

For the data-access code, it can be placed outside enclave, while imposing affordable security overhead. Concretely, one can build an authenticated data structure (ADS) on the untrusted data and attach a $O(\log{N})$ proof to individual read/write operations for their authenticity.

Running the code outside enclave leads to more OCalls (one Put/Get, at least one OCall) than that of running the code inside enclave. For the former, one Put/Get operation causes at least one OCall, while for the latter, it causes a OCall only when it flushes or misses a read buffer (which can be amortized to multiple Put/Get operations). Due to this reason, we think the alternative design by placing code outside enclave is also unfavorable from the performance perspective.\footnote{Building an LSM tree outside enclave also require non-trivial engineering efforts implementing the digest structure.}

\ifdefined\TTUT
\section{More on Data Confidentiality}
\label{sec:confidential2}

\subsubsection{Motivating Application}

{\bf Big-data outsourcing to the cloud}. Suppose a Twitter-alike startup company wants to outsource the storage of tweets to a third-party cloud (e.g., for economic reasons). The cloud storage service stores outsourced tweet database and serves two types of user operations: 1) data writes or updates as social-network users post new tweets, 2) data reads when a user reads her friends' tweets or when the startup company wants to analyze the tweets stream in real time for mining trendy stories (e.g., the ``Moments'' page in the Twitter website).

\subsection{Data Confidentiality by Deterministic Encryption}
\providecommand{\seckey}{{\relax\ifmmode{}sk\else $sk$\fi}\xspace}

In \elsm, data confidentiality is handled by adding an encryption layer on top. With a symmetric encryption, \elsm encrypts data keys when receiving \lkvPut/\lkvGet requests from the application enclave and decrypt data when returning results (i.e., data value in \lkvGet) to the applications. That is,

\begin{eqnarray}
&& \lkvPut(\enc_{\seckey}^{\de}(\datakey),\enc_{\seckey}^{\probenc}(\dataval))
\nonumber \\
\langle{}\datakey,\dataval,\ts_q\rangle{} &=&\dec_\seckey^{\probenc}(\lkvGet(\enc_{\seckey}^{\de}(\datakey),\ts_q))
\end{eqnarray}

Here, the symmetric key, \seckey, is used for both encryption and decryption. It is securely stored in the enclave. In addition, we use a deterministic encryption scheme (DE)~\cite{DBLP:conf/crypto/BellareBO07,DBLP:books/crc/KatzLindell2007} to support the exact-match query (in \lkvGet). Because DE has weaker security than the standard encryption, we limit DE to be used for encrypting data keys. The data value in a key-value record is encrypted using the standard probabilistic encryption (e.g., AES) that is semantically secure~\cite{DBLP:books/crc/KatzLindell2007}.

Concretely, in \elsmtwo, the trusted application deterministic-encrypts a data key and sends the ciphertext of the data key in the \lkvPut/\lkvGet requests to \elsm. It then conducts the buffer lookups based on the data-key ciphertext. In the case of buffer miss, it finds the target level, file name, and file block based on the data-key ciphertext.

In \elsmone, it relies on the native file-encryption provided by SGX SDK (see Figure~\ref{fig:systemstack}). The data blocks in the encrypted file are looked up based on the cleartext of data key (unlike in \elsmtwo).

In both designs, the files store the ciphertext of key-value records in a sorted order. The access sequence, that is, the indices of the data blocks being accessed, is disclosed to the untrusted host, which may reveal the relative order of data keys being accessed. In general, this work does not address the access-pattern side-channel security, for which existing works~\cite{DBLP:journals/iacr/SasyGF17,DBLP:conf/ndss/AhmadKSL18} can provide complementary supports.

In our implementation, we used the Intel SDK function \texttt{SGX\_RIJNDAEL128GCM\_ENCRYPT}~\cite{me:sgxsdk} with an empty IV (initialization vector) to encrypt each record. We have tested that in this setting, the repeated encryption (of the same plaintext and secret key) produces the same ciphertext. Thus we believe the use of an empty IV gives us a deterministic encryption scheme.

{\bf Confidentiality in range query}: To support range queries with confidentiality, we use Order-Preserving Encryption (OPE)~\cite{DBLP:conf/sp/PopaLZ13} where the ciphertext preserves the ordering of plaintext data. By this means, the enclave can search the OPE ciphertext of the queried range  over the OPE ciphertext of the data records buffered in the untrusted memory. In other words, the enclave runs the same range-query code in the vanilla (and unsecured) key-value store.

\fi





\end{document}